\documentclass[aps, A4paper, 12pt]{article}
\usepackage{slashed}
\usepackage{graphics}
\usepackage{epsfig}
\usepackage{subfigure}
\usepackage{graphics}
\usepackage{amsmath}
\usepackage{amsfonts}
\usepackage{amssymb}
\usepackage{url}
\usepackage{hyperref}
\newcommand{\be}{\begin{equation}}
\newcommand{\ee}{\end{equation}}
\newcommand{\bea}{\begin{eqnarray}}
\newcommand{\eea}{\end{eqnarray}}

\def\t{\hat t}
\def\s{\hat s}
\def\u{\hat u}
\begin{document}
\begin{titlepage}
\pagestyle{plain}
\begin{flushright}
KEK-TH-1202\\
December 29, 2007
\end{flushright}
\vspace{4\baselineskip}
\begin{center}
{\Large\bf 
Unparticle physics at the photon collider
}
\end{center}
\vspace{1cm}
\begin{center}
{\large
Tatsuru Kikuchi$^a$
\footnote{\tt E-mail:tatsuru@post.kek.jp},
Nobuchika Okada$^{a, b}$
\footnote{\tt E-mail:okadan@post.kek.jp}
and Michihisa Takeuchi$^{a, b, c}$
\footnote{\tt E-mail:tmichihi@post.kek.jp}
}
\end{center}
\vspace{0.2cm}
\begin{center}
${}^{a}$ 
{\small \it Theory Division, KEK,
Oho 1-1, Tsukuba, Ibaraki, 305-0801, Japan}\\
${}^{b}$ {\small \it Department of Particle and Nuclear Physics,
The Graduate University for Advanced Studies, \\
Oho 1-1, Tsukuba, Ibaraki, 305-0801, Japan}\\
${}^{c}$ {\small \it Yukawa Institute for Theoretical Physics, Kyoto University, Kyoto, 606-8502, Japan}\\
\vskip 10mm
\end{center}
\vskip 10mm
\begin{abstract}
Recently, a conceptually new physics beyond the Standard Model (SM), 
 unparticle, has been proposed, 
 where a hidden conformal sector is coupled to the SM sector 
 through higher dimensional operators. 
In this setup, we investigate unparticle physics 
 at the photon collider, in particular, 
 unparticle effects on the $\gamma \gamma \to \gamma \gamma$ process. 
Since this process occurs at loop level in the SM, 
 the unparticle effects can be significant 
 even if the cutoff scale is very high. 
In fact, we find that the unparticle effects cause sizable deviations 
 from the SM results. 
The scaling dimension of the unparticle $d_{\cal U}$ 
 reflects the dependence of the cross section 
 on the final state photon invariant mass, 
 so that precision measurements of this dependence 
 may reveal the scaling dimension of the unparticle. 

\end{abstract}
\end{titlepage}
\section{Introduction}
The Large Hadron Collider (LHC), which will start its operation within a year, 
is expected to probe a new hitherto unexplored domain of particles 
and forces beyond the standard model around TeV scale. 
Although the LHC has the considerable potential to detect 
 some indication of new physics beyond the Standard Model (SM), 
 the detailed study of its properties needs more precise measurements 
 and such a work will be performed 
 at the International Linear Collider (ILC). 
According to the ILC Reference Design Report \cite{ILC}, 
 the ILC is determined to run with $\sqrt{s} = 500$ GeV 
 and the total luminosity required is ${\cal L} = 500~ {\rm fb}^{-1}$ 
 within the first four years of operation 
 and ${\cal L} = 1000~ {\rm fb}^{-1}$ 
 during the first phase of operation with $\sqrt{s} = 500$ GeV.
An $e^+ e^-$ collider is uniquely capable of operation at a series 
 of energies near the threshold of a new physics process. 
This is an extremely powerful tool for precision measurements of 
 particle masses and unambiguous particle spin determinations. 
Various ILC physics studies indicate that 
 a $\sqrt{s} = 500$ GeV collider can have a great impact 
 on understanding a new physics around TeV scale. 
An energy upgrade up to $\sqrt{s} \sim 1$ TeV would open 
 the door to even greater discoveries. 

Another very unique feature of the ILC is that it can be transformed 
 into $\gamma \gamma$ collisions with the photon beams 
 generated by using the Compton backscattering of the initial electron 
 and laser beams. 
In this case, the energy and luminosity of the photon beams would be 
 the same order of magnitude of the original electron beams. 
Since the set of final states at a photon collider 
 is much richer than that in the $e^+ e^-$ mode, 
 the photon collider would open a wider window 
 to probe new physics beyond the SM. 
In fact, it has been seen in several new physics models 
 that photon collider is more powerful for searching models 
 than the $e^+ e^-$ linear collider.

The most comprehensive description of the photon collider 
 available at present is in a part of the TESLA TDR \cite{TESLA}. 
Also, there are some useful reviews for the physics 
 at the photon collider as an option of the ILC \cite{Bechtel, Telnov}. 
Since the high energy photon beams are provided 
 through Compton scatterings from the electron beams, 
 the $\gamma \gamma$ luminosity is determined by the geometric 
 luminosity of the original electron beams \cite{Ginzburg:1981vm}. 
For the standard ILC beam parameters, 
 the $\gamma \gamma$ luminosity is expected to be 
 $L_{\gamma \gamma} = 0.17 \times L_{ee}$ 
 with the integrated luminosity of the incident $e^+e^-$ collider 
 ($L_{ee}$). 
Considering that cross sections in ${\gamma \gamma}$ 
 are larger than those in $e^+ e^-$ collisions 
 by one order of magnitude, the number of events 
 will be somewhat larger than in $e^+ e^-$ collisions.

A certain class of new physics models includes a scalar field 
 which is singlet under the SM gauge group.  
Such a new particle can have a direct coupling with photons 
 suppressed by a new physics scale in low energy effective theory. 
If the new physics scale is low enough, the particle can be 
 produced at the photon collider, and thus the photon collider 
 can be a powerful tool to probe such a class of new physics models. 
In particular, the process, $\gamma \gamma \to \gamma \gamma$,  
 is interesting because in the SM, this process occurs only 
 at loop level and the SM background for new physics search 
 is expected to be small.

As one of such models, in this paper, we consider 
 a new physics recently proposed by Georgi \cite{Georgi:2007ek}, 
 which is described in terms of 'unparticle'. 
The unparticle physics is originated from a theory 
 having some conformal fixed points in low energy, 
 and the interaction between this conformal hidden sector 
 and the SM sector is described by the effective theory at low energy.
A concrete example of unparticle staff was proposed 
 by Banks-Zaks \cite{Banks:1981nn} many years ago, 
 where providing a suitable number of massless fermions, 
 theory reaches a non-trivial infrared fixed points 
 and a conformal theory can be realized at a low energy. 
Various phenomenological considerations on the unparticle physics 
 have recently been developed in the literature \cite{U-pheno, U-pheno2} 
 as well as some studies on the formal aspects 
 of the unparticle physics \cite{U-formal}. 
There have also been studied on the astrophysical and cosmological 
 applications of the unparticle physics \cite{U-astro, Kikuchi2}, 
 especially in \cite{Kikuchi2}, even the possibility 
 for the unparticle to be a dark matter has been proposed. 

In this paper, we investigate the unparticle physics 
 at the photon collider. 
We concentrate on the process, $\gamma \gamma \to \gamma \gamma$, 
 and the unparticle effects on it. 
As mentioned above, there is no tree level contribution in the SM, 
 and we find that the unparticle effects cause sizable deviations 
 from the SM results even if the cutoff scale of the higher 
 dimensional interaction is of order 10 TeV.

\section{Basics of unparticle physics}
We begin with a brief review of the basic structure of the unparticle physics. 
First we introduce a coupling between the new physics operator 
 ($\cal{O}_{\rm UV}$) with dimension $d_{\rm UV}$ 
 and the Standard Model one (${\cal O}_{\rm SM}$) 
 with dimension $n$ at some ultraviolet (UV) scale as
\bea
 {\cal L} = \frac{c_n}{M^{d_{\rm UV}+n-4}} 
     \cal{O}_{\rm UV} {\cal O}_{\rm SM} ,  
\eea
where $c_n$ is a dimension-less constant, and $M$ is the energy scale 
 characterizing the new physics. 
This new physics sector is assumed to become conformal 
 at an IR scale $\Lambda_{\cal U}$, and 
 the operator $\cal{O}_{\rm UV}$ flows to the unparticle operator 
 ${\cal U}$ with dimension $d_{\cal U}$. 
In low energy effective theory, we have the operator of the form, 
\bea
{\cal L}=c_n 
 \frac{\Lambda_{\cal U}^{d_{\rm UV} - d_{\cal U}}}{M^{d_{\rm UV}+n-4}}   
 {\cal U} {\cal O}_{\rm SM} 
\equiv 
  \frac{\lambda_n}{\Lambda^{d_{\cal U}+ n -4}}  {\cal U} {\cal O}_{\rm SM},  
\eea 
where the dimension of the unparticle ${\cal U}$ has been 
 matched by $\Lambda_{\cal U}$ which is induced by 
 the dimensional transmutation, 
 $\lambda_n$ is an order one coupling constant
 and $\Lambda$ is the (effective) cutoff scale of 
 low energy effective theory. 
In this paper, we consider only the scalar unparticle. 

It was found in Ref.~\cite{Georgi:2007ek}
 that, by exploiting scale invariance of the unparticle, 
 the phase space for an unparticle operator 
 with the scaling dimension $d_{\cal U}$ and momentum $p$ 
 is the same as the phase space for 
 $d_{\cal U}$ invisible massless particles, 
\begin{eqnarray}
d \Phi_{\cal U}(p) = 
 A_{d_{\cal U}} \theta(p^0) \theta(p^2)(p^2)^{d_{\cal U}-2} 
 \frac{d^4p}{(2\pi)^4} \,,
\label{Phi}
\end{eqnarray}
where
\begin{eqnarray}
A_{d_{\cal U}} = \frac{16 \pi^{\frac{5}{2}}}{(2\pi)^{2 d_{\cal U}}}
\frac{\Gamma(d_{\cal U}+\frac{1}{2})}{\Gamma(d_{\cal U}-1) 
\Gamma(2 d_{\cal U})}.
\label{A}
\end{eqnarray}
Also, based on the argument on the scale invariance, 
 the (scalar) propagator for the unparticle was suggested 
 to be \cite{U-pheno}: 
\bea
{\cal P}(P^2) 
=
\left\{
\begin{array}{ll}
Z_{d_{\cal U}} \times |P^2|^{d_{\cal U}-2} \;~~~(P^2 < 0) \;, \\
Z_{d_{\cal U}} \times e^{-i \pi d_{\cal U}} |P^2|^{d_{\cal U}-2} ~~(P^2 > 0) \;. 
\end{array}
\right. 
\label{propagator}
\eea
where 
 $Z_{d_{\cal U}} \equiv \frac{A_{d_{\cal U}}}{2\sin(\pi d_{\cal U})}$ 
 with $Z_{d_{\cal U} \to 1} \to -1$.
Interestingly, $d_{\cal U}$ is not necessarily integer, 
 it can be any real number or even complex number. 
In this paper we consider the scaling dimension 
 in the range,  $1 \le  d_{\cal U} <  2$,  for simplicity.

For our study on the photon collider, 
 we consider the interaction between the unparticle and photons of the form
\footnote{
When we introduce all those kinds of terms 
 between the unparticle and the SM gauge bosons, 
 the process $ gg \to {\cal U} \to \gamma \gamma$ 
 has an impact on physics at hadron colliders 
 such as the LHC and Tevatron. 
In particular, there is an impact on Higgs boson ($h$) search 
 through the gluon fusion process, $ gg \to h \to  \gamma \gamma$. 
Although such a process is out of our scope in this paper, 
 it is worth investigating. 
}: 
\bea
{\cal L}_{\rm int} = 
  \frac{\cal U}{\Lambda^{d_{\cal U}}} F_{\mu \nu} F^{\mu \nu} \;.
\label{intU-gamma}
\eea
This interaction causes the process $\gamma \gamma \to \gamma \gamma$ 
 mediated by the unparticle in the $s$, $t$ and $u$-channels 
 at the tree level.

\section{Unparticle effects at the Photon Collider}
Now we consider the effects of unparticle on 
 the $\gamma\gamma \to \gamma \gamma$ process 
 at the photon colliders. 
The helicity amplitude for the process
\be
\gamma (p_1,\lambda_1) \gamma (p_2,\lambda_2) \to
\gamma (p_3,\lambda_3) \gamma (p_4,\lambda_4) \ \ ,
\ee
is denoted as ${\cal M}_{\lambda_1 \lambda_2 \lambda_3 \lambda_4}(\s,\t,\u)$
\footnote{
We will use the notation for the matrix elements 
for the photon photon scattering amplitude as
$\left< \gamma (p_3,\lambda_3) \gamma (p_4,\lambda_4)  | \gamma (p_1,\lambda_1) \gamma (p_2,\lambda_2) \right> 
= 1 + i (2 \pi)^4 \delta^4 (p_1 +  p_2 - p_3 -p_4) 
{\cal M}_{\lambda_1 \lambda_2 \lambda_3 \lambda_4}$.}, 
where 
$\s=(p_1+p_2)^2$, 
$\t=(p_3-p_1)^2$,
$\u=(p_4-p_1)^2$.
The Bose-Einstein statistics demands 
\bea
{\cal M}_{\lambda_1 \lambda_2 \lambda_3\lambda_4}(\s,\t,\u) &=&
{\cal M}_{\lambda_2 \lambda_1 \lambda_4\lambda_3}(\s,\t,\u) \ ,
\label{Bose2} \\
{\cal M}_{\lambda_1 \lambda_2 \lambda_3\lambda_4}(\s,\t,\u) &=&
{\cal M}_{\lambda_2 \lambda_1 \lambda_3\lambda_4}(\s,\u,\t) \ ,
\label{Bose1}
\eea
while  crossing symmetry implies
\bea
{\cal M}_{\lambda_1 \lambda_2 \lambda_3\lambda_4}(\s,\t,\u) =&
{\cal M}_{-\lambda_4 \lambda_2 \lambda_3 -\lambda_1}(\t,\s,\u) =&
{\cal M}_{\lambda_1 -\lambda_3 -\lambda_2\lambda_4}(\t,\s,\u) \ ,
\label{cross-ts}\\
{\cal M}_{\lambda_1 \lambda_2 \lambda_3\lambda_4}(\s,\t,\u) =&
{\cal M}_{-\lambda_3 \lambda_2 -\lambda_1\lambda_4}(\u,\t,\s) =&
{\cal M}_{\lambda_1 -\lambda_4 \lambda_3 -\lambda_2}(\u,\t,\s) \ .
\label{cross-us}
\eea
When parity and time inversion invariance holds, 
 we have, respectively, the constraints 
\bea
{\cal M}_{\lambda_1 \lambda_2 \lambda_3\lambda_4}(\s,\t,\u) &=&
{\cal M}_{-\lambda_1-\lambda_2- \lambda_3-\lambda_4}(\s,\t,\u) \ \ ,
 \label{parity} \\
{\cal M}_{\lambda_3 \lambda_4 \lambda_1\lambda_2}(\s,\t,\u) & = &
{\cal M}_{\lambda_1 \lambda_2 \lambda_3\lambda_4}(\s,\t,\u) \ \  .
\label{time-inv}
\eea
As a result, the 16 possible helicity amplitudes can 
 be expressed in terms of only three independent amplitudes, 
${\cal M}_{++++}(\s,\t,\u)$, ${\cal M}_{+++-}(\s,\t,\u)$ and 
${\cal M}_{++--}(\s,\t,\u)$, through the relations \cite{Jikia:1993tc}, 
\bea
 {\cal M}_{\pm\pm\mp\pm}(\s,\t,\u)&= &{\cal M}_{\pm\mp\pm\pm}(\s,\t,\u)=
{\cal M}_{\pm\mp\mp\mp}(\s,\t,\u)= {\cal M}_{\pm\pm\pm\mp}(\s,\t,\u), \label{+++-} \\
{\cal M}_{--++}(\s,\t,\u)& = &{\cal M}_{++--}(\s,\t,\u) \ , \label{++--} \\
 {\cal M}_{\pm\mp\pm\mp}(\s,\t, \u)&=&{\cal M}_{----}(\u,\t,\s)=
{\cal M}_{++++}(\u,\t,\s) \ , \label{++++1} \\
 {\cal M}_{\pm\mp\mp\pm}(\s,\t, \u)&=&{\cal M}_{\pm\mp\pm\mp}(\s,\u, \t)=
{\cal M}_{++++}(\t,\s, \u)= {\cal M}_{++++}(\t,\u, \s) \ .
\label{++++2}
\eea
Hence all the combinations can be expressed in terms of only three quantities,
${\cal M}_{++++}$, ${\cal M}_{++--}$ and ${\cal M}_{+++-}$.

The resultant helicity amplitudes in the SM, $ {\cal M}^{\rm SM}_{\lambda_1 \lambda_2 \lambda_3\lambda_4}$, 
are well-studied, for example, in \cite{Jikia:1993tc, Gounaris:1999gh}.
In the numerical calculation, we make use 
 of {\tt LoopTools} \cite{looptools} for evaluating the loop functions
 that appear in the SM background.

\begin{figure}[p]
\begin{center}
\centering
\epsfxsize=4truein
\hspace*{0in}
\rotatebox{-90}{
\epsffile{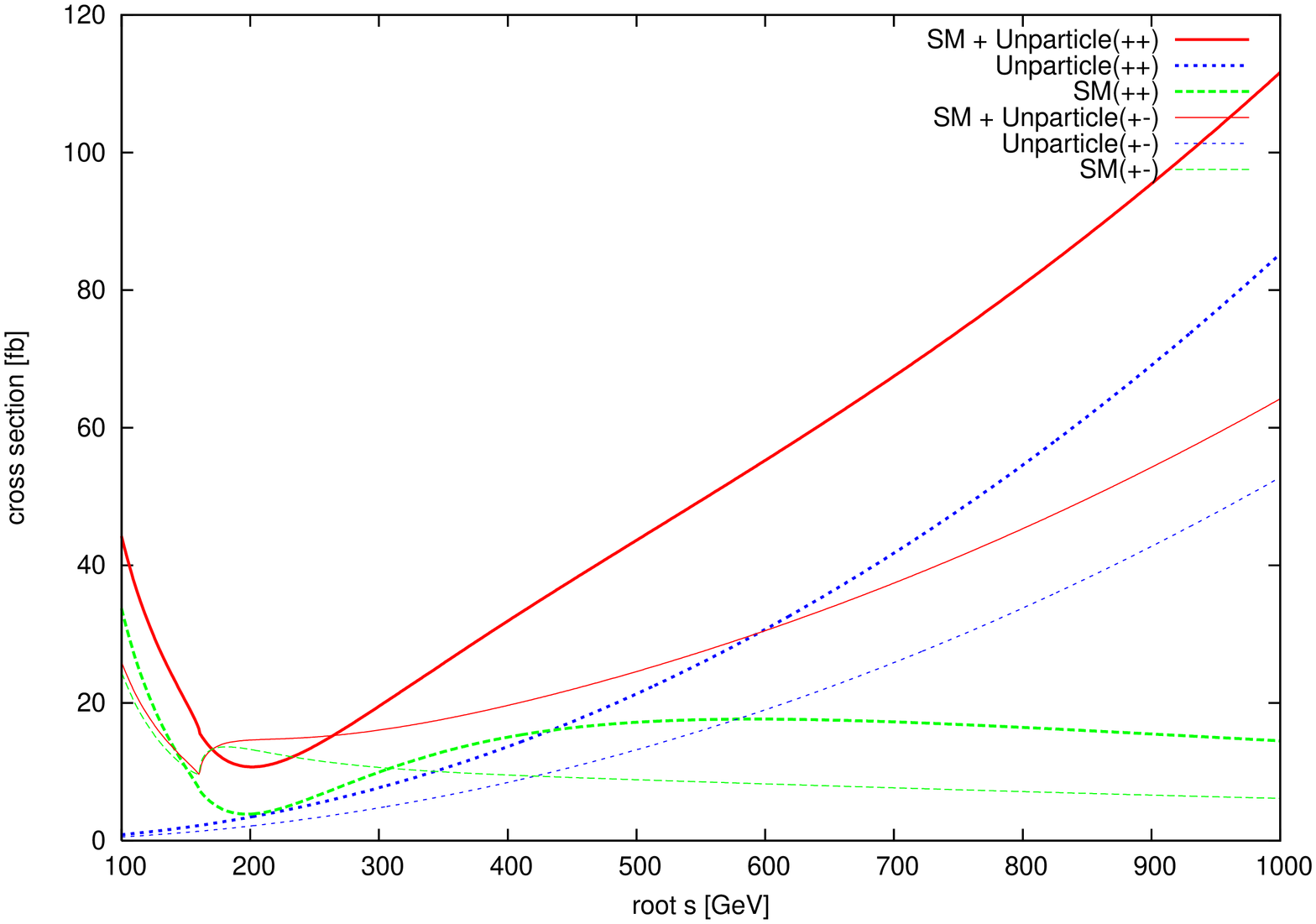}}
\end{center}
\caption{
The total cross section of the scattering, 
 $\gamma \gamma \to \gamma \gamma$, 
 for the Standard Model process, purely unparticle contribution, 
 and the combined result as a function of the photon energy $\sqrt{\s}$.
Here we have taken the limit $d_{\cal U} \to 1$ 
 and the cutoff scale to be $\Lambda=5$ TeV. 
In the integration with respect to the scattering angle, 
 we have imposed a cut as  $ 30^\circ < \theta < 150^\circ$. 
Two possible combinations of the initial photon helicities 
 $(\lambda_1 \lambda_2)=(++),~(+-)$ are taken into account 
 in this analysis, and the results are shown by different thickness 
 of each lines.  
}
\label{Fig1}
\end{figure}

\begin{figure}[p]
\begin{center}
\centering
\epsfxsize=4truein
\hspace*{0in}
\rotatebox{-90}{
\epsffile{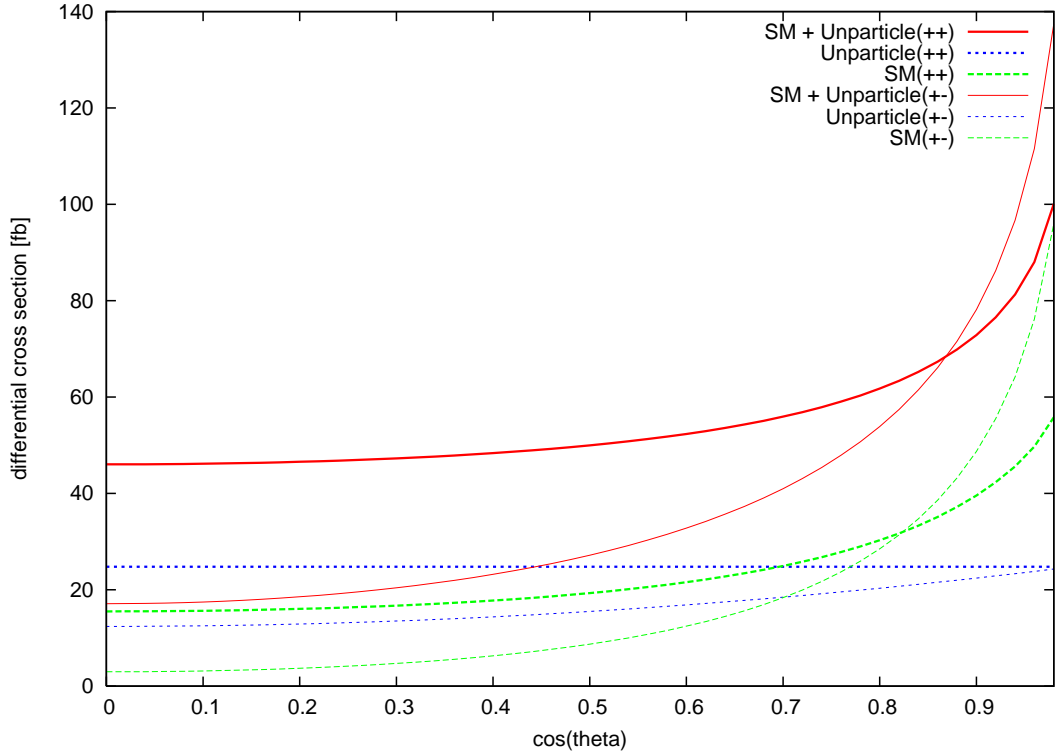}}
\end{center}
\caption{
The angular distribution for 
 the $\gamma \gamma \to \gamma \gamma$ process 
 with the initial photon energy $\sqrt{\s} = 500$ GeV 
 and the cutoff scale $\Lambda=5$ TeV, 
 in the limit of $d_{\cal U} \to 1$. 
}
\label{Fig2}
\end{figure}

\begin{figure}[p]
\begin{center}
\centering
\epsfxsize=4truein
\hspace*{0in}
\rotatebox{-90}{
\epsffile{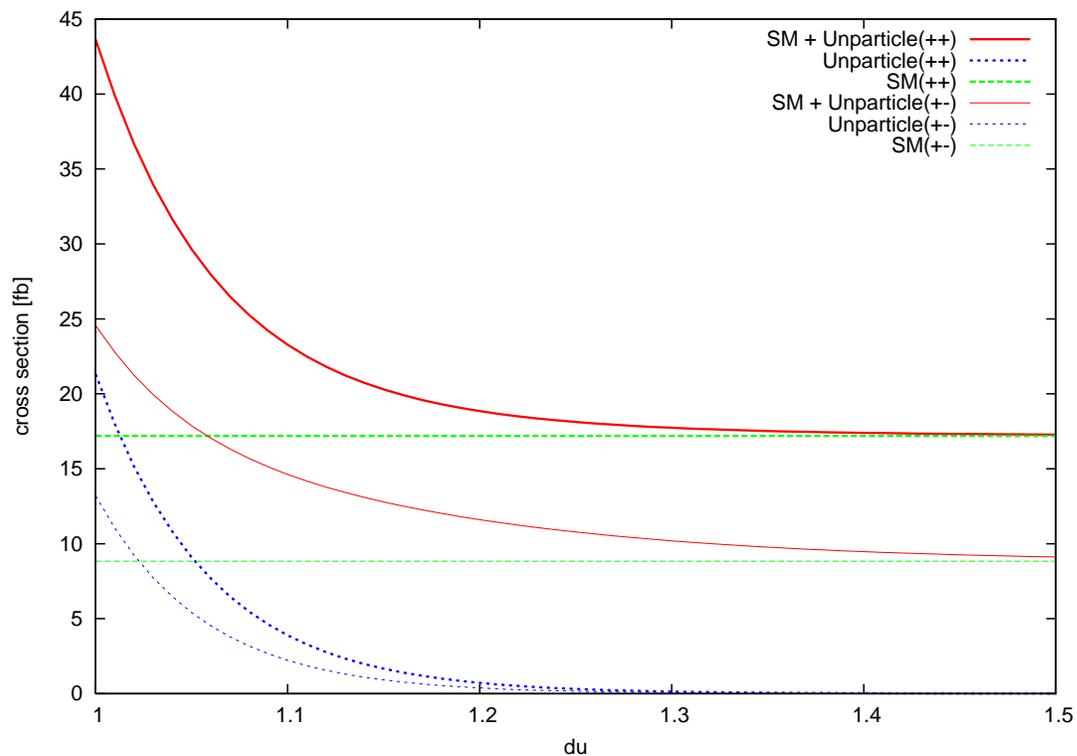}}
\end{center}
\caption{
The total cross section of the $\gamma \gamma \to \gamma \gamma$ process 
 for the Standard Model contribution, the pure unparticle contribution, 
 and the combined result as a function of the scaling dimension,
 $d_{\cal U}$, 
 for the initial photon energy $\sqrt{\s} = 500$ GeV 
 and $\Lambda=5$ TeV.  
We have again imposed a cut for the scattering angle 
 as $ 30^\circ < \theta < 150^\circ$ in the integration. 
}
\label{Fig3}
\end{figure}

It is easy to calculate the helicity amplitudes 
 for the $\gamma \gamma \to \gamma \gamma$ process 
 mediated by the unparticle in the $s$, $t$ and $u$-channels: 
\begin{enumerate}
\item $s$-channel
\bea
i {\cal M}^{{\cal U} (s)}_{\lambda_1 \lambda_2 \lambda_3\lambda_4} 
=  - \frac{4 \, \s^2}{\Lambda^{2 d_{\cal U}}} 
 {\cal P}(\s) \delta_{\lambda_1, \lambda_2} \delta_{\lambda_3, \lambda_4} \:.
\eea
\item $t$-channel
\bea
i {\cal M}^{{\cal U} (t)}_{\lambda_1 \lambda_2 \lambda_3\lambda_4} 
= - \frac{4 \, \t^2}{\Lambda^{2 d_{\cal U}}} {\cal P}(\t) 
\delta_{\lambda_1, -\lambda_3} \delta_{\lambda_2, -\lambda_4} \:.
\eea
\item $u$-channel
%
\bea
i {\cal M}^{{\cal U} (u)}_{\lambda_1 \lambda_2 \lambda_3\lambda_4}
= - \frac{4 \, \u^2}{\Lambda^{2 d_{\cal U}}} {\cal P}(\u) 
\delta_{\lambda_1, -\lambda_3} \delta_{\lambda_2, -\lambda_4} \:.
\eea
\end{enumerate}
%
%
The differential polarized cross section with respect to the scattering angle $\theta$ is given by
\be
\frac{d \hat{\sigma}^{(\lambda_1 \lambda_2)}}{d \cos \theta} = 
\frac{1}{ 32 \pi \s} \sum_{\lambda_3, \lambda_4} 
\left[\left| {\cal M}^{\rm SM}_{\lambda_1 \lambda_2 \lambda_3\lambda_4}  
+  {\cal M}^{{\cal U} (s)}_{\lambda_1 \lambda_2 \lambda_3\lambda_4}  
+  {\cal M}^{{\cal U} (t)}_{\lambda_1 \lambda_2 \lambda_3\lambda_4} 
+  {\cal M}^{{\cal U} (u)}_{\lambda_1 \lambda_2 \lambda_3\lambda_4} \right|^2 \right] \;.
\ee
The resultant cross sections in the limit $d_{\cal U} \to 1$,
 as a function of the photon beam energy are shown in
 Fig.~\ref{Fig1}.  
Here we have taken the cutoff scale to be $\Lambda=5$ TeV.  
Contributions of the unparticle mediated processes become 
 dominant as the beam energy becomes larger, as expected. 
The angular distribution of the cross section for $d_{\cal U} \to 1$ 
with a fixed photon beam energy, $\sqrt{\s} = 500$ GeV, 
is depicted in Fig.~\ref{Fig2}. 
The SM cross sections have a peak in the forward (and backward) region, 
 while the cross sections of the unparticle mediated processes  
 are almost flat, reflecting the 0-spin of the scalar unparticle. 
Fig.~\ref{Fig3} shows the resultant cross section 
 as a function of the scaling dimension $d_{\cal U}$, 
 for a fixed photon beam energy $\sqrt{\s} = 500$ GeV  
 and the cutoff scale $\Lambda=5$ TeV.
The unparticle effects quickly go down 
 as $d_{\cal U}$ becomes larger, 
 as expected in the results of the helicity amplitudes 
 for the unparticle mediated processes. 

In order to obtain the realistic cross section 
 $\sigma(\gamma \gamma \to \gamma \gamma)$ 
 at the photon collider, 
 we convolute the fundamental cross section 
 $\hat{\sigma}(\gamma \gamma \to \gamma \gamma)$ 
 with the photon luminosity function.
Throughout this paper, $\sqrt{s}$ refers 
 to the center-of-mass energy of the incident $e^{+}e^{-}$ collider  
 and $\sqrt{\hat{s}}$ refers to the center of mass energy 
 of the two incoming photons. 
The laser backscattering \cite{TESLA, Ginzburg:1981vm} is the standard
and efficient technique to convert an electron beam into a photon beam.
The photon luminosity function 
 $F_{\gamma/e}(x)$ is given by \cite{TESLA, Ginzburg:1981vm}:
\bea
F_{\gamma/e}(P_e , P_\ell, x) &=& \frac1{D(\xi)} \left[\frac{1}{1-x} - x +(2 r-1)^2
- P_e \, P_\ell  \, \xi \, r (2 r -1) (2-x) \right] \;,
\eea
where $P_e$ is the polarization of the initial electron and $P_\ell$ is the degree of circular
polarization of the initial {\cal laser} beam ($|P_{e}| \leq1,~|P_{\ell}| \leq 1$), $r=x/\xi(1-x)$,
and $D(\xi)$ is a normalization factor,
\be
D(\xi)=\left(1-\frac4{\xi}-\frac{8}{\xi^2} \right)
\ln(1+\xi)+\frac12+\frac8\xi-\frac1{2(1+\xi)^2} \;,
\ee
with
\be
\xi = \frac{4 E_0 \omega_0}{m_e^2}  = 
4.475 \left(\frac{2 E_0}{500~{\rm GeV}} \right) \left(\frac{\omega_0}{1.17~{\rm eV}} \right) \;,
\ee
where $E_0$ is the energy of the incident electron and $\omega_0$ 
 is the energy of the incident laser photon.
The Compton kinematics are characterized by the variable $x$, 
 and one finds maximal energy conversion is given at $x_{\rm max} = \xi/(\xi+1) = 0.817$
 for $D(\xi = 4.475) \simeq 1.77$.
Then the maximum photon energy is given by 
 $\omega_{\rm max} = x_{\rm max} E_0 \simeq 0.82 E_0$. 
This means that about $82 \%$ of the incident electron-positron 
 beam energy can be transmitted into the photon collider.
One of the important observation is that the spectrum depends on the product of the
helicity of the electron and the laser beam. The backscattered photons will retain a certain
amount of the polarization of the laser photon beam.
The hardest spectrum is provided by choosing the circular polarization of the laser ($P_\ell$) 
and the mean helicity of the electron ($P_e$) to be opposite, $P_e P_\ell = -1$, for both arms of the collider. 
$F_{\gamma/e}(0 , 0 , x)$ corresponds to the unpolarized case.

Photon beam helicity is given by \cite{TESLA, Ginzburg:1981vm}:
\bea
\langle h_{\gamma} (x) \rangle 
&\equiv& \frac{- P_{\ell} \, (2 r-1) \left[ 1/(1-x) + 1-x  \right]
+ P_e \, \xi \, r \left[1 + (1 - x) (2 r-1)^2 \right]}{D(\xi) \; F_{\gamma/e}(P_e , P_\ell,x)} \;.
\eea
By using the photon beam helicity, the total photon luminosity function can be decomposed 
according to each helicity component as
\be
F^{\pm}_{\gamma/e}(x, P_e, P_\ell) 
\equiv \frac{1\pm \langle h_{\gamma} (x) \rangle }{2} F_{\gamma/e}(P_e , P_\ell,x)  \;.
\ee
There is a relation between different sign choices of polarization vectors:
\be
F^{\lambda}_{\gamma/e}(x, P_e, P_\ell )  = F^{-\lambda}_{\gamma/e}(x, -P_e,-P_\ell)  \;.
\label{relation}
\ee
Then, the cross section for the polarized photon beam can be obtained by integrating over all the energy distributions:
\bea
&&\sigma(\gamma\gamma\to \gamma\gamma)
\cr
&& \ \ \ \ \ \ \ \ =\sum_{\lambda_1,\lambda_2} \int_{x_{1 {\rm min}}}^{x_{\rm max}} \!\int_{x_{2 {\rm min}}}^{x_{\rm max}} 
F^{\lambda_1}_{\gamma/e}(x_1, P_e, P_\ell ) F^{\lambda_2}_{\gamma/e}(x_2, P_e^\prime, P_\ell^\prime )\,
\hat{\sigma}^{(\lambda_1 \lambda_2)}(\gamma\gamma\to \gamma\gamma; \;
\hat{s}=x_1 x_2 s) dx_1 dx_2 
\nonumber\\
&& \ \ \ \ \ \ \ \ =\sum_{\lambda_1,\lambda_2}
\int_{\tau_{\rm min}}^{\tau_{\rm max}} \frac{d{\cal L}^{(\lambda_1 \lambda_2)}}{d\tau} (\tau, P_e, P_\ell , P_e^\prime, P_\ell^\prime) \,
\hat{\sigma}^{(\lambda_1 \lambda_2)}(\gamma\gamma\to \gamma\gamma; \s = \tau s) d\tau \;, 
\label{conv}
\eea
where the maximum value of $x$ is given by $x_{\rm max} = \xi/(1+\xi) = 0.817$.
In the second line in the above, we made a change of variable, 
the corresponding maximal value of $\tau$ is $\tau_{\rm max} = x_{\rm max}^2 = 0.668$.
In our numerical evaluation of Eq.~(\ref{conv}), we introduce an infrared cutoff 
$\tau_{\rm min} = 0.04$ ($\sqrt{s}>100$ GeV) which is necessary to avoid the infrared divergence
in the cross section that mainly comes from the fermion loop contributions.

The luminosity function in the above formula is defined as follows:
\bea
\frac{d{\cal L}^{(\lambda_1 \lambda_2)}}{d\tau} (\tau, P_e, P_\ell , P_e^\prime, P_\ell^\prime ) 
&\equiv&  
 \int_{x_{1 {\rm min}}}^{x_{\rm max}} \int_{x_{2 {\rm min}}}^{x_{\rm max}}  
 F_{\gamma/e}^{\lambda_1}(x_1, P_e, P_\ell ) F_{\gamma/e}^{\lambda_2}(x_2, P_e^\prime, P_\ell^\prime ) \delta(\tau - x_1 x_2) \; dx_1 dx_2 
\nonumber\\
&=&
\int_{-y_{\rm max}}^{y_{\rm max}} 
F_{\gamma/e}^{\lambda_1}(\sqrt{\tau} e^y , P_e, P_\ell ) F_{\gamma/e}^{\lambda_2}(\sqrt{\tau} e^{-y} , P_e^\prime, P_\ell^\prime ) dy  \;,
\label{Luminosity}
\eea
where $y_{\rm max} \equiv \frac{1}{2} \log \frac{\tau_{\rm max}}{\tau}$.

Based on the relation of Eq.~(\ref{relation}) and the exchange symmetry between 
$F_{\gamma/e}^{\lambda_1}$ and $F_{\gamma/e}^{\lambda_2}$ in the definition of 
the luminosity function, Eq.~(\ref{Luminosity}), it holds the following symmetry property:
\bea
{\cal L}^{+-} (P_e, P_\ell ,P_e^\prime, P_\ell^\prime ) 
&=&
{\cal L}^{++} (P_e, P_\ell ,-P_e^\prime,-P_\ell^\prime ) \;, 
\nonumber\\
{\cal L}^{-+} (P_e,P_\ell , P_e^\prime, P_\ell^\prime )
&=&
{\cal L}^{++} (-P_e,-P_\ell , P_e^\prime, P_\ell^\prime ) \;, 
\nonumber\\
{\cal L}^{--} (P_e, P_\ell , P_e^\prime, P_\ell^\prime )
&=&
{\cal L}^{++} (-P_e^\prime,-P_\ell^\prime ,-P_e,-P_\ell ) \;.
\eea
Hence, without loss of generality, we can consider only 
${\cal L}^{++}(P_e, P_\ell , P_e^\prime, P_\ell^\prime )$.
(Here, we denote $\frac{d{\cal L}^{\lambda_1 \lambda_2}}{{d\tau}}(\tau)$ as ${\cal L}^{\lambda_1 \lambda_2}$, for simplicity.)

\begin{figure}[p]
\begin{center}
\subfigure[Each $(\lambda_1,\lambda_2)=(\pm\pm),(\pm\mp)$ contribution]{\includegraphics[angle=-90, width=0.7\textwidth]{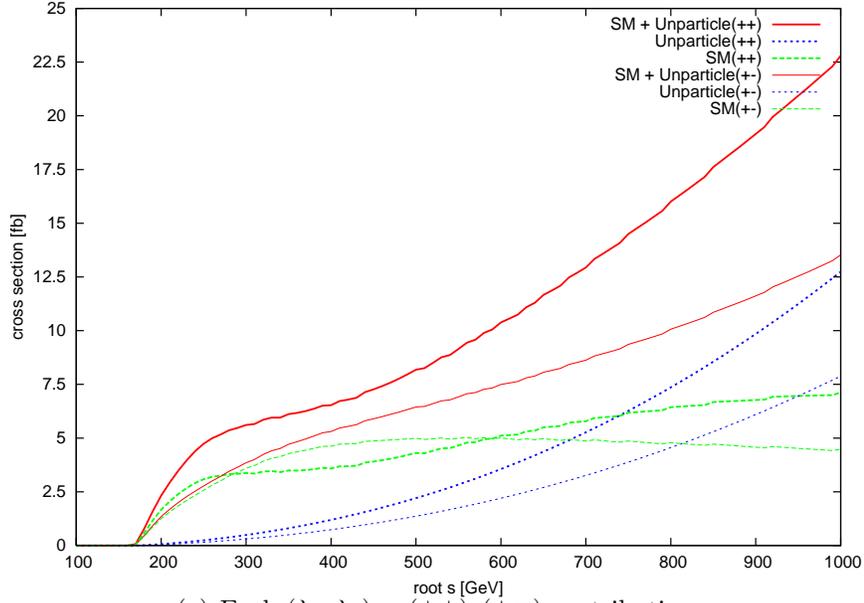}}
\subfigure[Sum of both contributions]{\includegraphics[angle=-90, width=0.7\textwidth]{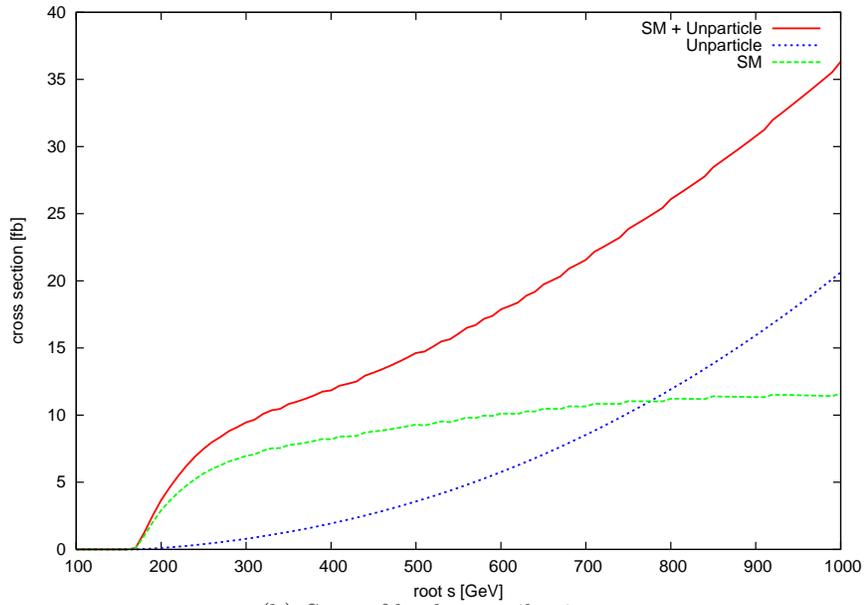}}
\end{center}
\caption{
The convoluted cross section of 
 the $\gamma \gamma \to \gamma \gamma$ process 
in the case of unpolarized beams.
The figure shows for the Standard Model case, pure unparticle case
 and the combined result as a function of the incident 
 $e^+ e^-$ collider energy $\sqrt{s}$.
The top figure shows each contribution for $(\lambda_1,\lambda_2)=(\pm\pm)$ 
and $(\pm\mp)$.
The bottom figure shows the sum of both contributions.}
\label{Fig4}
\end{figure}

\begin{figure}[p]
\begin{center}
\subfigure[Each $(\lambda_1,\lambda_2)=(\pm\pm),(\pm\mp)$ contribution]{\includegraphics[angle=-90, width=0.7\textwidth]{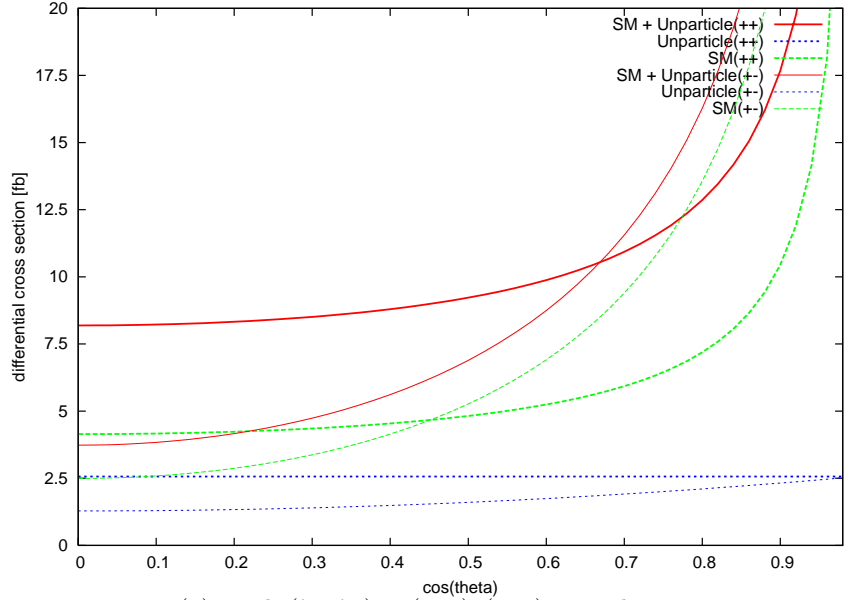}}
\subfigure[Sum of both contributions]{\includegraphics[angle=-90, width=0.7\textwidth]{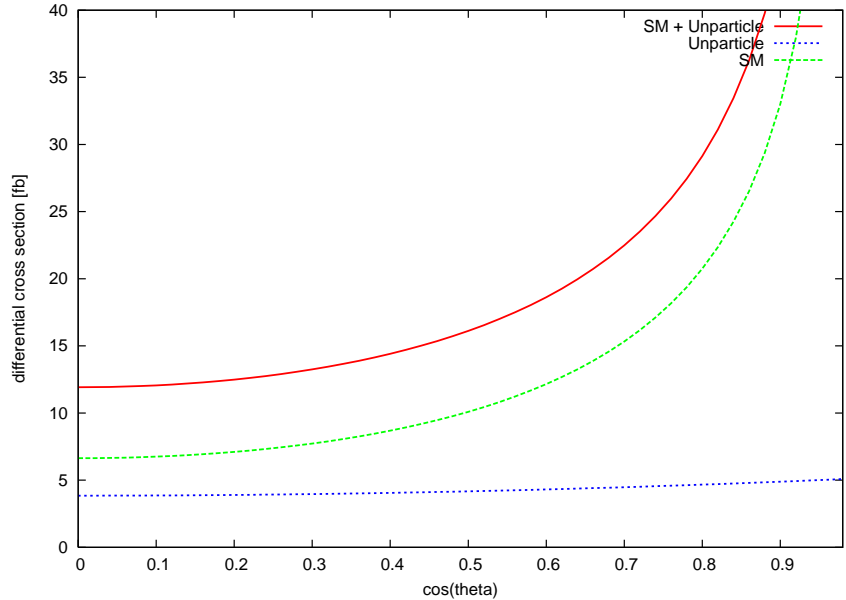}}
\end{center}
\caption{
The angular distribution for the process, $\gamma \gamma \to \gamma \gamma$,
 including the unparticle in the intermediate state. 
In this figure, we have fixed the incident $e^+ e^-$ beam energy as $\sqrt{s} = 500$ GeV.
The top figure shows each contribution for $(\lambda_1,\lambda_2)=(\pm\pm)$ 
and $(\pm\mp)$.
The bottom figure shows the sum of both contributions.}
\label{Fig5}
\end{figure}

\begin{figure}[p]
\begin{center}
\subfigure[Each $(\lambda_1,\lambda_2)=(\pm\pm),(\pm\mp)$ contribution]{\includegraphics[angle=-90, width=0.7\textwidth]{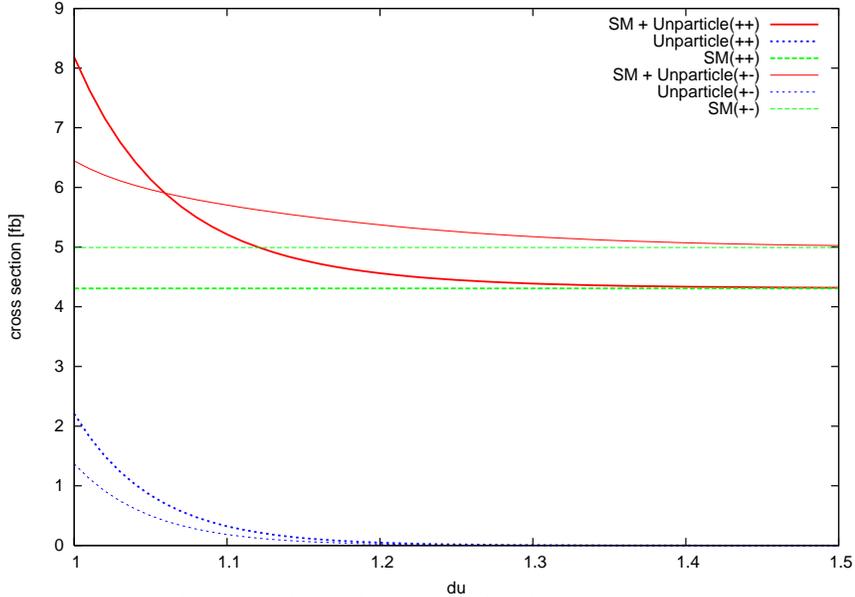}}
\subfigure[Sum of both contributions]{\includegraphics[angle=-90, width=0.7\textwidth]{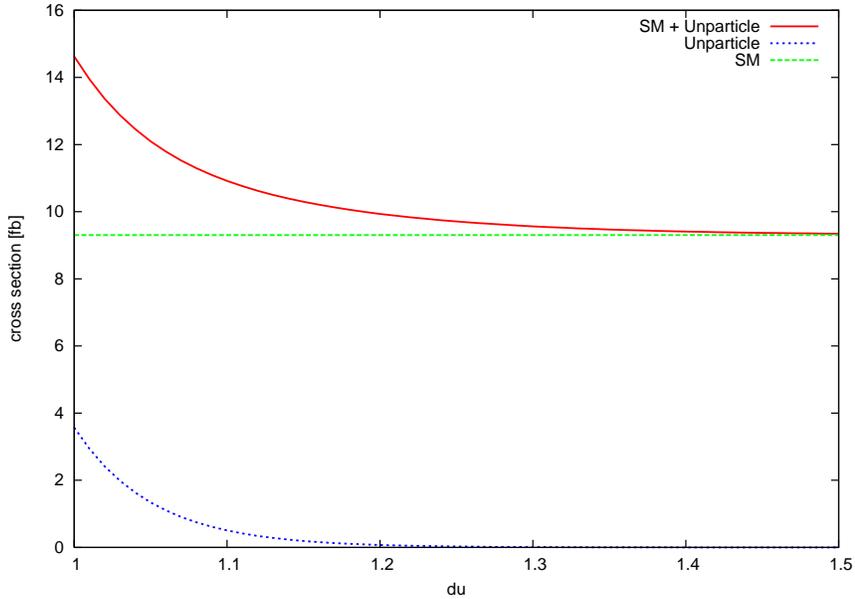}}
\end{center}
\caption{
The total cross section of 
the $\gamma \gamma \to \gamma \gamma$ process 
for the Standard Model contribution, 
pure unparticle contribution
and the combined result as a function of the scaling dimension, $d_{\cal U}$.
We have fixed the incident $e^+e^-$ beam energy as $\sqrt{s} = 500$ GeV
and $\Lambda=5$ TeV.
A cut for the scattering angle is taken to be $ 30^\circ < \theta < 150^\circ$.
The top figure shows each contribution for $(\lambda_1,\lambda_2)=(\pm\pm)$ 
and $(\pm\mp)$.
The bottom figure shows the sum of both contributions.
}
\label{Fig6}
\end{figure}

Corresponding to Fig.~{\ref{Fig1}}, Fig.~{\ref{Fig2}} and Fig.~{\ref{Fig3}},
 the results after the convolution  are shown in Fig.~{\ref{Fig4}}, Fig.~{\ref{Fig5}} and Fig.~{\ref{Fig6}}.
In these figures, we have used the unpolarized luminosity function to show the results 
in case of unpolarized beam.
Fig.~{\ref{Fig4}} shows the total cross section ($ 30^\circ < \theta < 150^\circ$) 
as a function of the incident $e^+ e^-$ collider energy $\sqrt{s}$.
Fig.~{\ref{Fig5}} shows the angular distribution of the cross section. 
Fig.~{\ref{Fig6}} shows the total cross section ($ 30^\circ < \theta < 150^\circ$)
 for a fixed photon beam energy $\sqrt{s} = 500$ GeV
 as a function of the scaling dimension $d_{\cal U}$.
We can see that in the case with $d_{\cal U} \simeq 1$, 
there are sizable deviations from the SM results 
 for $\sqrt{s}=500$ GeV, for example, 
 even though the cutoff scale is very high $\Lambda=5$ TeV. 

\begin{figure}[p]
\begin{center}
\subfigure[$(\lambda_1,\lambda_2)=(\pm\pm)$ contributions]{\includegraphics[angle=-90, width=0.7\textwidth]{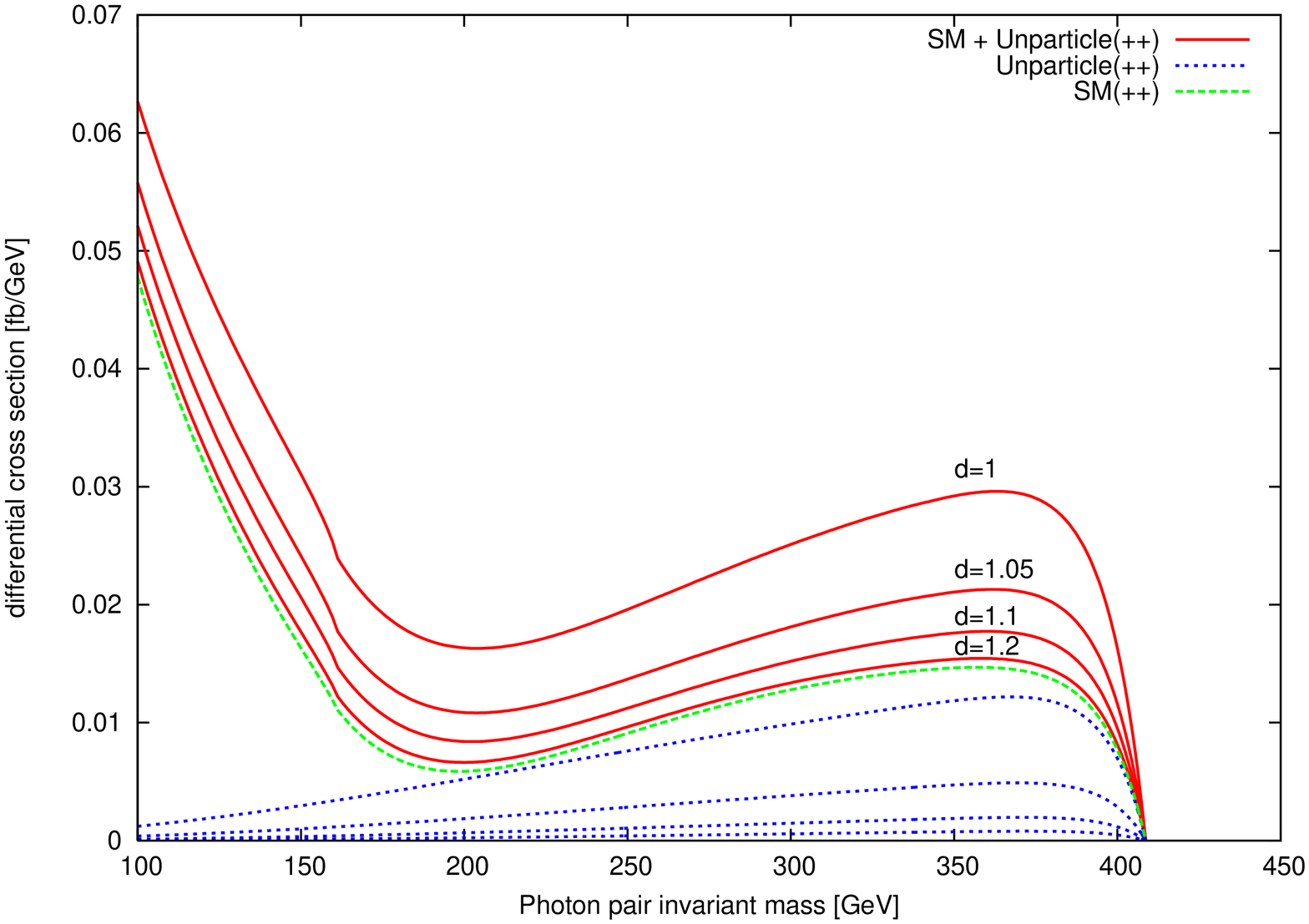}}
\subfigure[$(\lambda_1,\lambda_2)=(\pm\mp)$ contributions]{\includegraphics[angle=-90, width=0.7\textwidth]{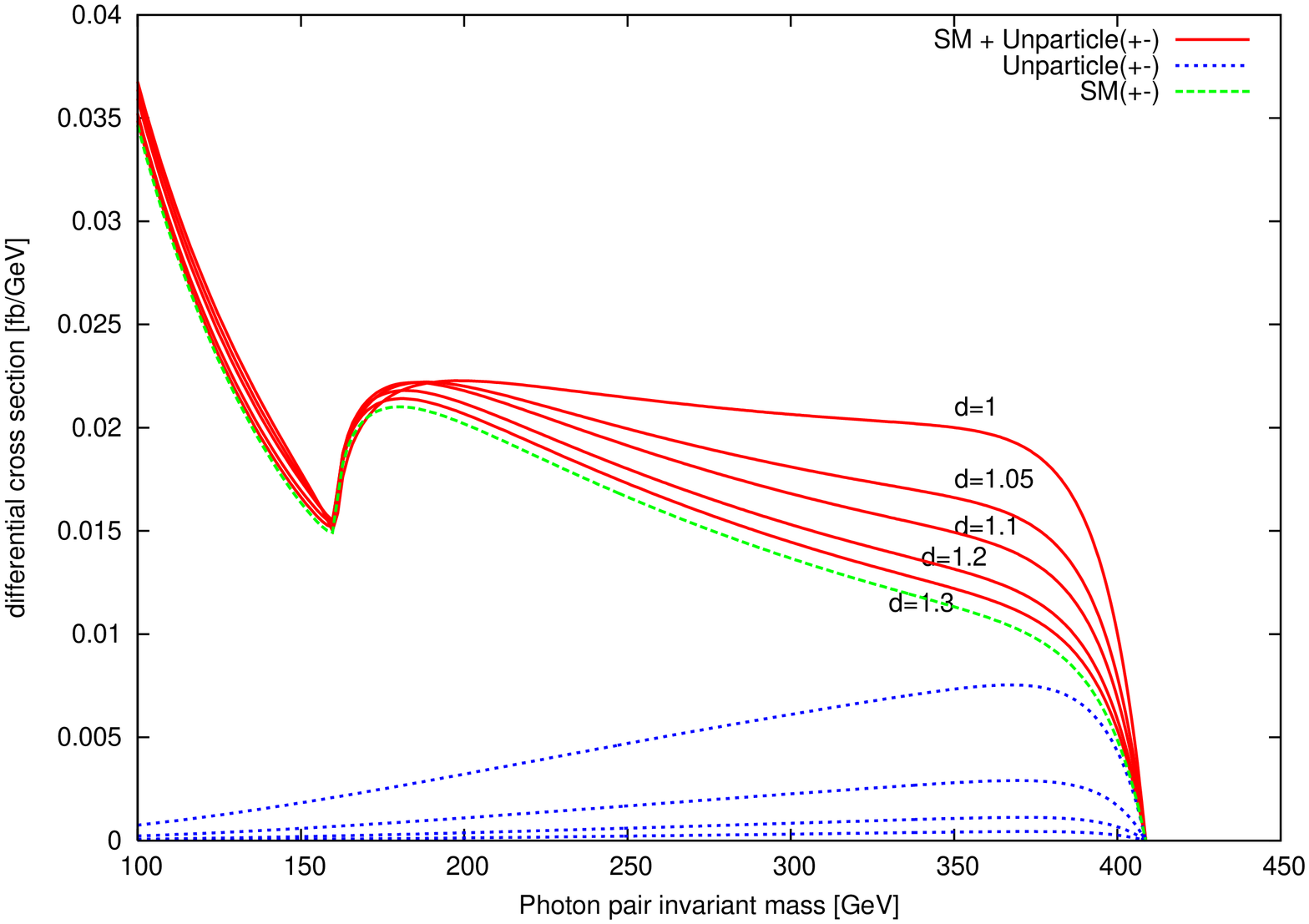}}
\end{center}
\caption{
The differential cross section $d\sigma/dM_{\gamma \gamma}$ 
for the process $\gamma \gamma \to  \gamma \gamma$ ($ 30^\circ < \theta < 150^\circ$)
as a function of a final state photon invariant mass $M_{\gamma \gamma}$.
The energy is fixed to $\sqrt{s} =500$ GeV and $\Lambda=5$ TeV.
Here we show each contribution from different incident photon beam helicity combination separately
The top figure is for $(\lambda_1,\lambda_2)=(\pm\pm)$ and the bottom is for $(\pm\mp)$.
Each curve corresponds to $d_{\cal U}=1, ~1.05,~1.1,~1.2,~1.3$. 
}
\label{Fig7-1}
\end{figure}

\begin{figure}[p]
\begin{center}
\subfigure[The differential cross section]{\includegraphics[angle=-90, width=0.7\textwidth]{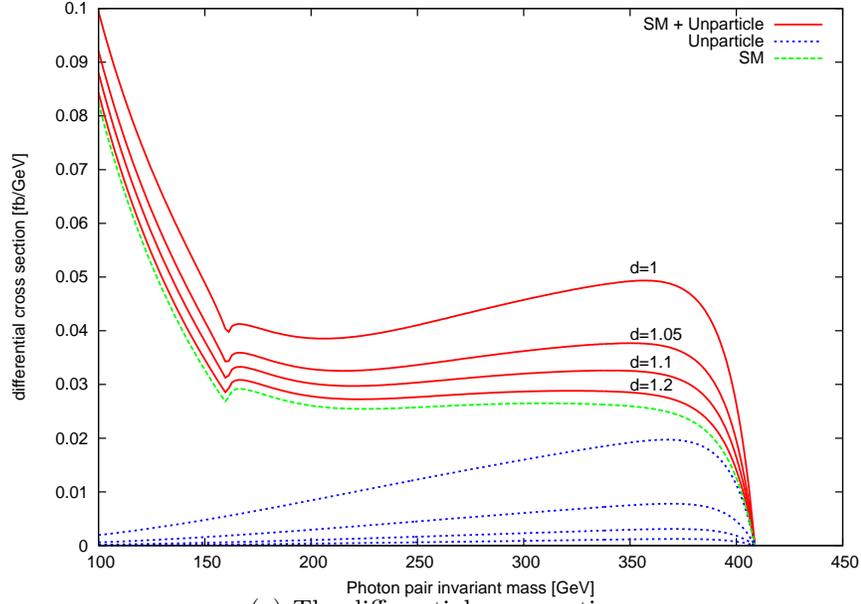}\label{7-2a}}
\subfigure[Ratio of cross section $\sigma_{{\cal U}+{\rm SM}}/\sigma_{\rm SM}$]
{\includegraphics[angle=-90, width=0.7\textwidth]{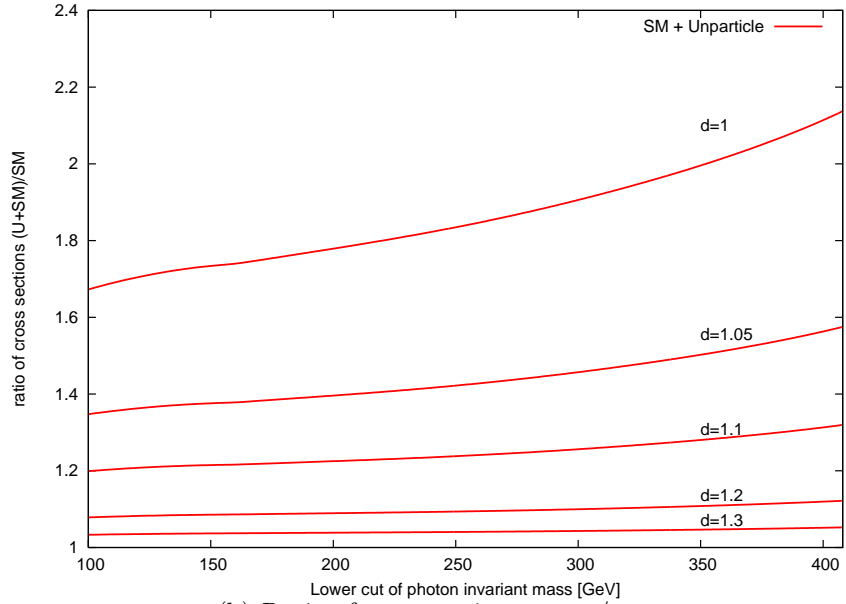}\label{7-2b}}
\end{center}
\caption{The total cross section for the case of unpolarized beams.
Fig.~\ref{7-2a} shows the same figure as Fig.~\ref{Fig7-1} but the initial helicities are summed up 
to show the unpolarized cross sections.
Fig.~\ref{7-2b} shows the ratio of the signal cross section to the SM cross section 
($\sigma_{{\cal U}+{\rm SM}}/\sigma_{\rm SM}$) 
 as a function of a lower energy cut on the final state photon invariant mass $M_{\gamma \gamma}^{\rm cut}$.
}
\label{Fig7-2}
\end{figure}

\begin{figure}[p]
\begin{center}
\subfigure[$(\lambda_1,\lambda_2)=(\pm\pm)$ contributions]{\includegraphics[angle=-90, width=0.7\textwidth]{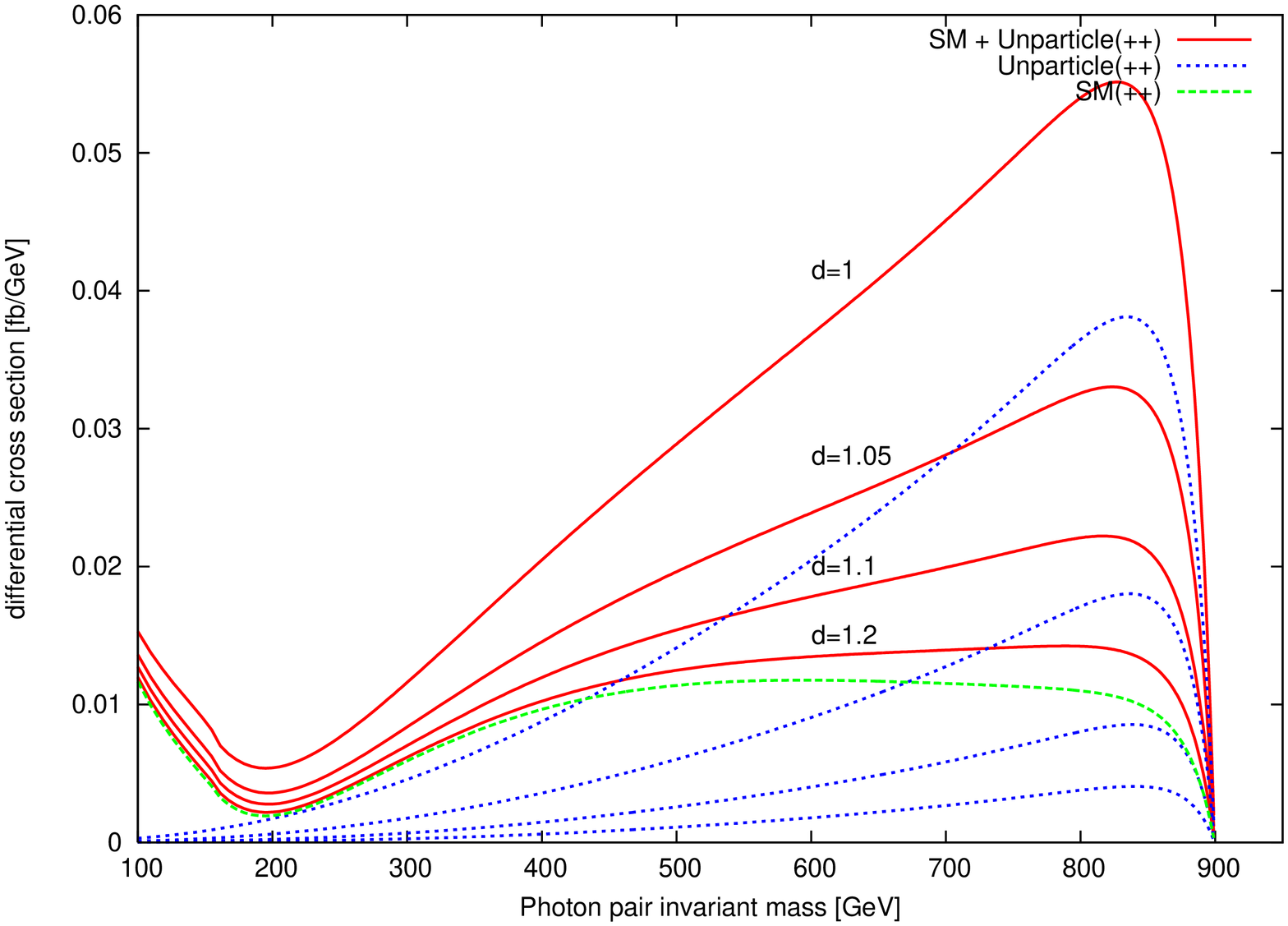}}
\subfigure[$(\lambda_1,\lambda_2)=(\pm\mp)$ contributions]{\includegraphics[angle=-90, width=0.7\textwidth]{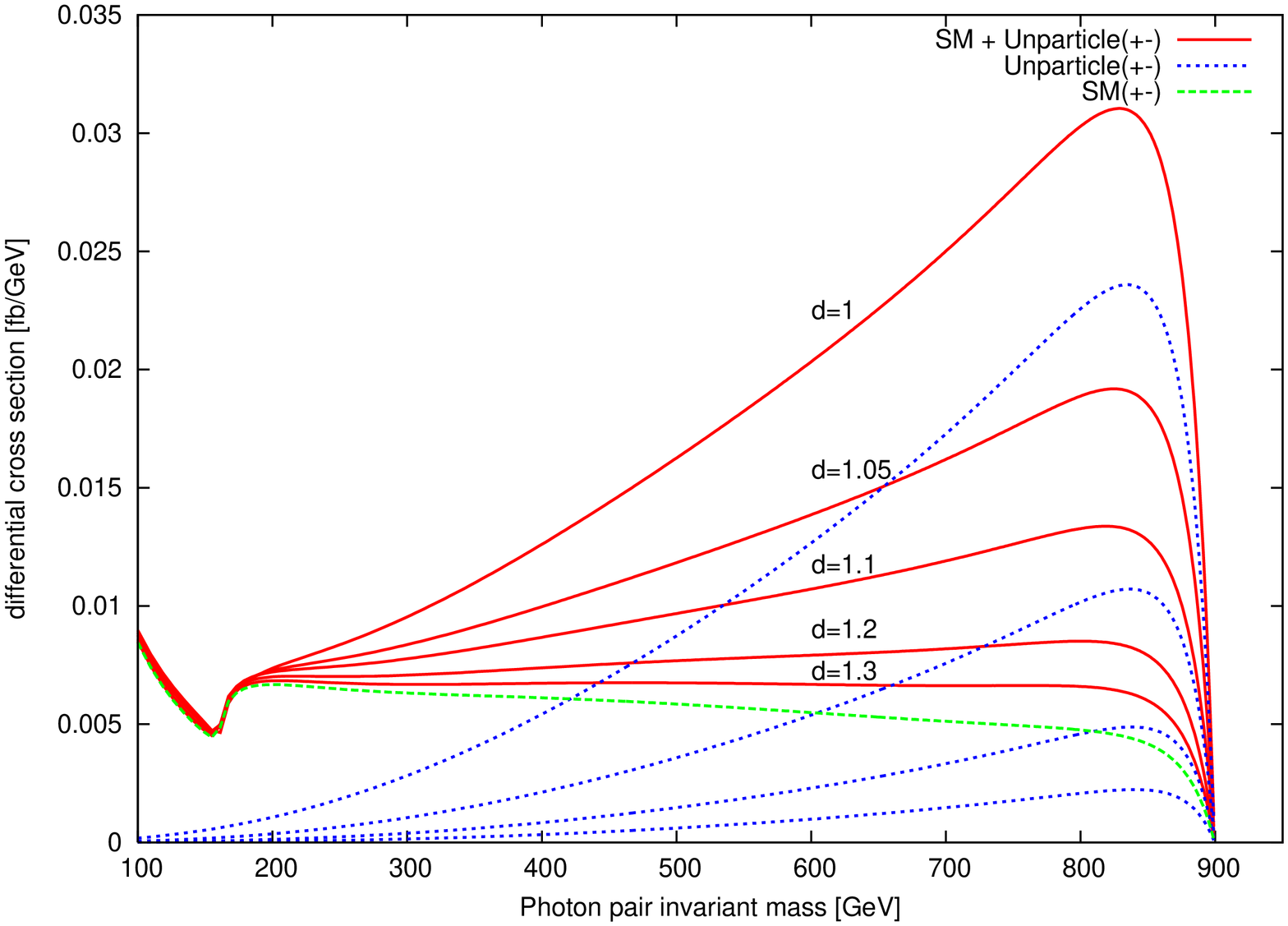}}
\end{center}
\caption{
The same figure as Fig.~\ref{Fig7-1}, but for the case of $\sqrt{s}=1$ TeV. 
}
 \label{Fig8-1}
\end{figure}

\begin{figure}[p]
\begin{center}
\subfigure[The differential cross section]{\includegraphics[angle=-90, width=0.7\textwidth]{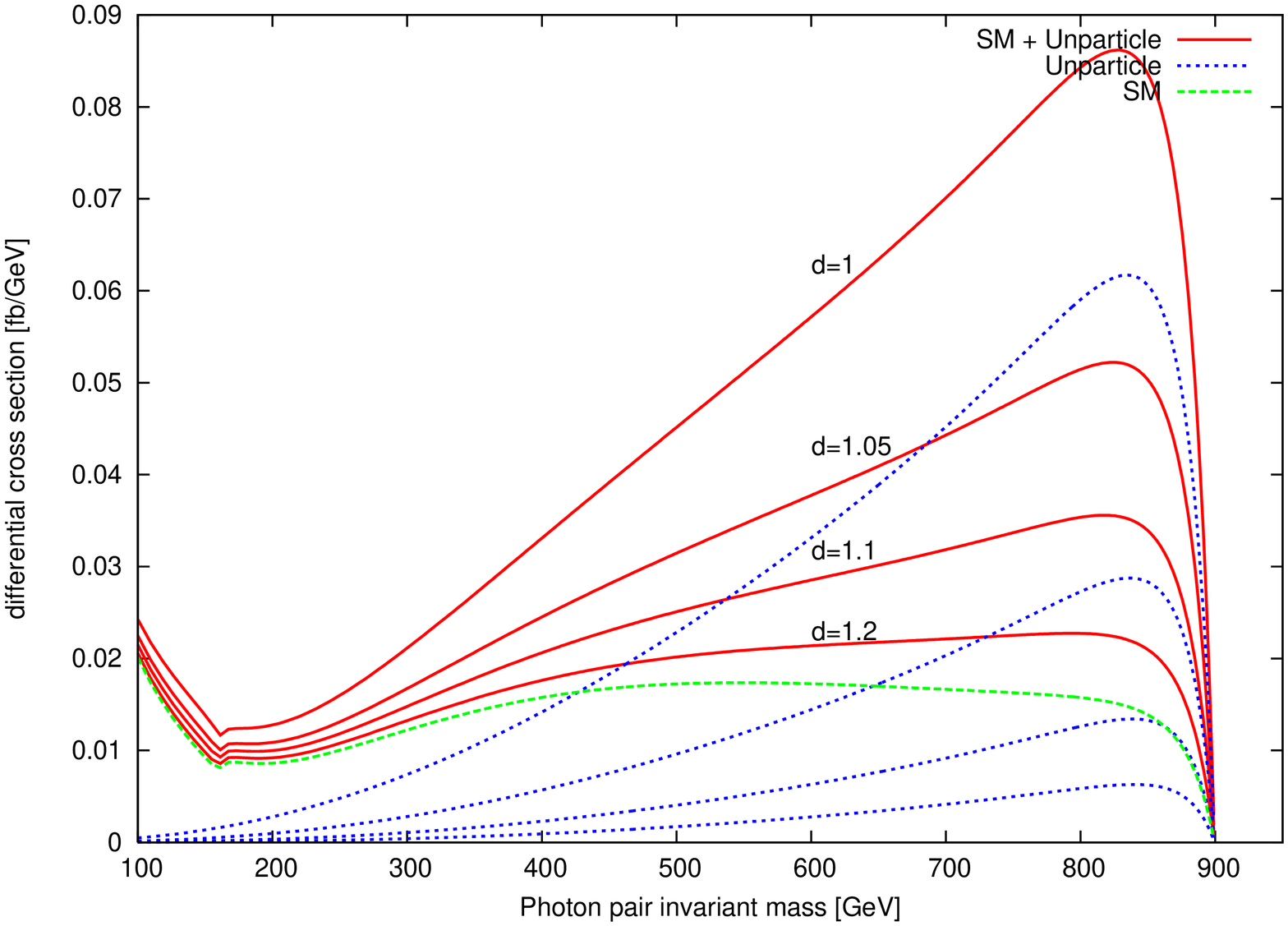}}
\subfigure[Ratio of cross section $\sigma_{{\cal U}+{\rm SM}}/\sigma_{\rm SM}$]{\includegraphics[angle=-90, width=0.7\textwidth]{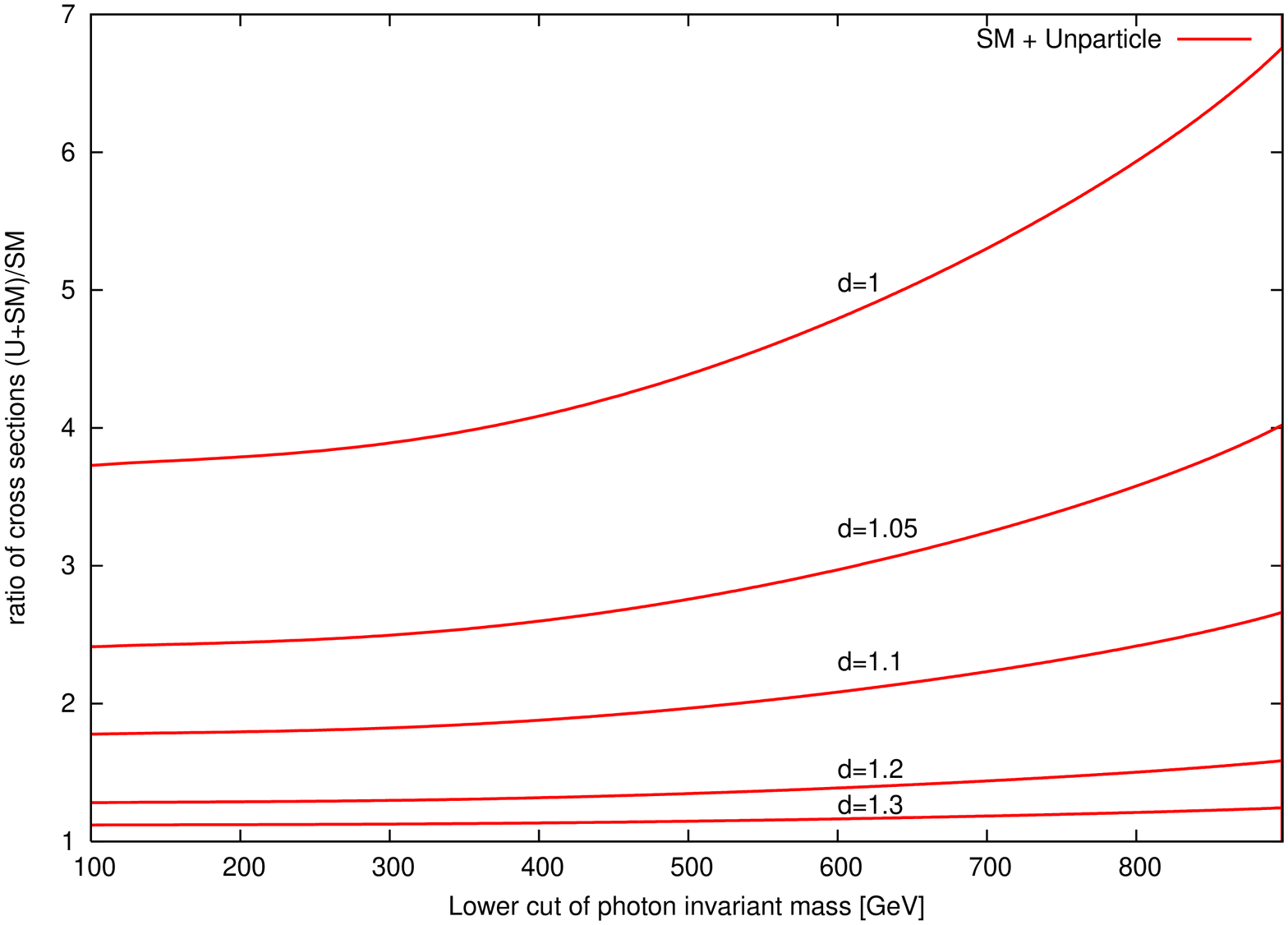}}
\end{center}
\caption{
The same figure as Fig.~\ref{Fig7-2}, but for the case of $\sqrt{s}=1$ TeV. 
}
\label{Fig8-2}
\end{figure}

In Fig.~\ref{Fig7-1}, we show the differential cross section as a function of  
 the invariant mass of the final state photons, 
$d\sigma/dM_{\gamma \gamma}$,
for $\sqrt{s}=500$ GeV and for various $d_{\cal U}$ values.
Here, we show the contributions from each helicity components $(\lambda_1 \lambda_2)=(++)$ and
from $(\lambda_1 \lambda_2)=(+-)$, separately.
However, only the sum of all these contributions is observable, 
and we show the result in Fig.~\ref{Fig7-2} (top).
The deviation from the SM result becomes larger as the invariant mass is raised. 
After imposing a lower energy cut for the photon invariant mass 
 ($M_{\gamma \gamma}^{\rm cut}$) and the integration with respect to 
 the invariant mass,
we obtain the cross section  as a function of 
 $M_{\gamma \gamma}^{\rm cut}$.
Fig.~\ref{Fig7-2} (bottom) shows the ratio of the signal cross section to the SM one as 
 a function of $M_{\gamma \gamma}^{\rm cut}$ 
 for $\sqrt{s}=500$ GeV, $\Lambda=5$ TeV and for various $d_{\cal U}$.
The ratio becomes enhanced for larger $M_{\gamma \gamma}^{\rm cut}$. 
The resultant cross sections show different behaviors 
as a function of $M_{\gamma \gamma}^{\rm cut}$, 
for different $d_{\cal U}$. 
Therefore we can determine $d_{\cal U}$ 
by precisely measuring the cross sections ratio
as a function of $M_{\gamma \gamma}^{\rm cut}$. 
The results for $\sqrt{s}=1$ TeV are also shown in Fig.~\ref{Fig8-1} and Fig.~\ref{Fig8-2}.

\begin{figure}[p]
\begin{center}
\subfigure[The differential cross section]{\includegraphics[angle=-90, width=0.7\textwidth]{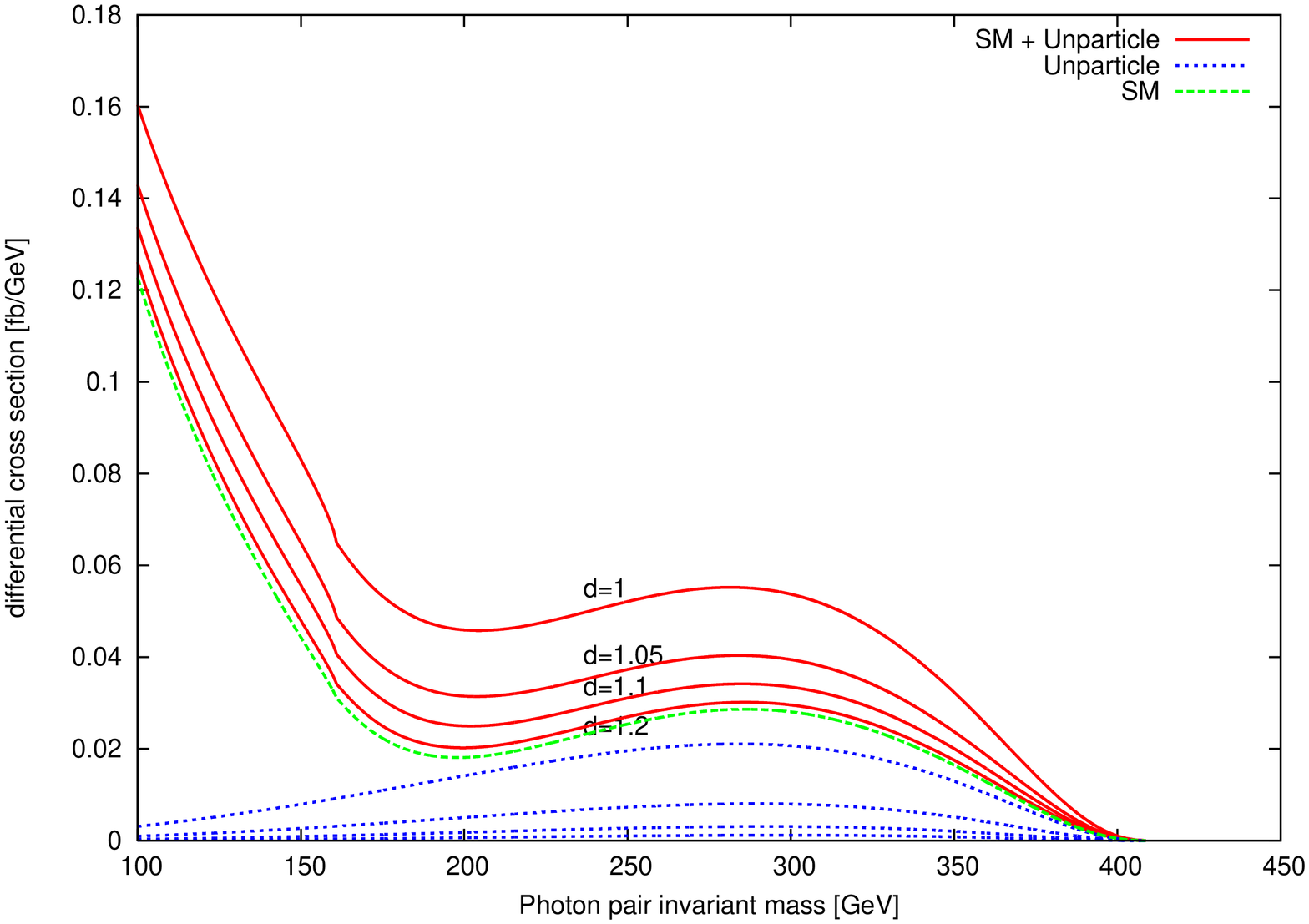}}
\subfigure[Ratio of cross section $\sigma_{{\cal U}+{\rm SM}}/\sigma_{\rm SM}$]{\includegraphics[angle=-90, width=0.7\textwidth]{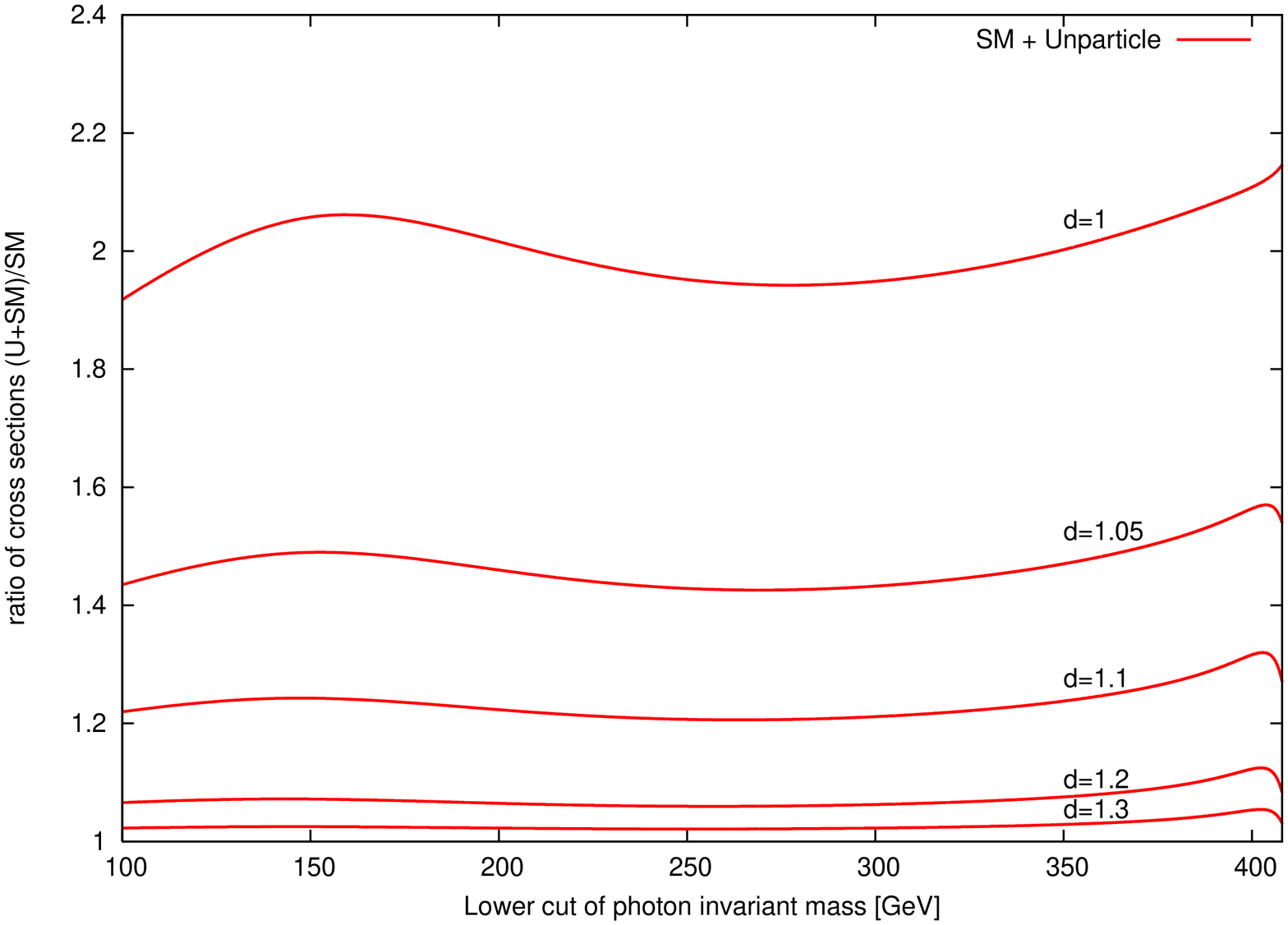}}
\end{center}
\caption{The same figure as Fig.~\ref{Fig7-2} 
but the beam polarizations are chosen as 
$(P_e, P_\ell,  P'_e, P'_\ell) = (\pm \pm \pm \pm)$.
}
\label{Fig9}
\end{figure}

\begin{figure}[p]
\begin{center}
\subfigure[The differential cross section]{\includegraphics[angle=-90, width=0.7\textwidth]{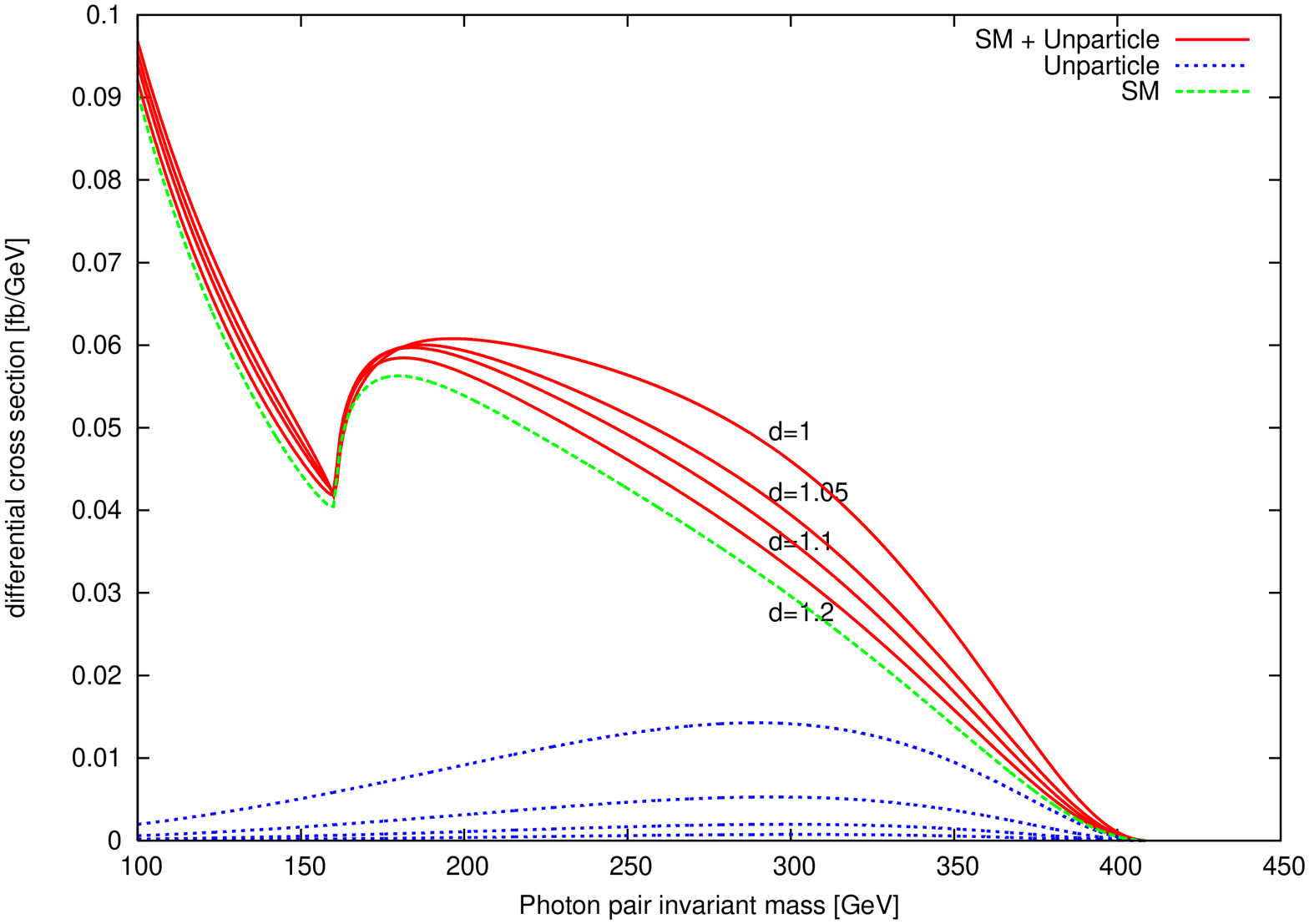}}
\subfigure[Ratio of cross section $\sigma_{{\cal U}+{\rm SM}}/\sigma_{\rm SM}$]{\includegraphics[angle=-90, width=0.7\textwidth]{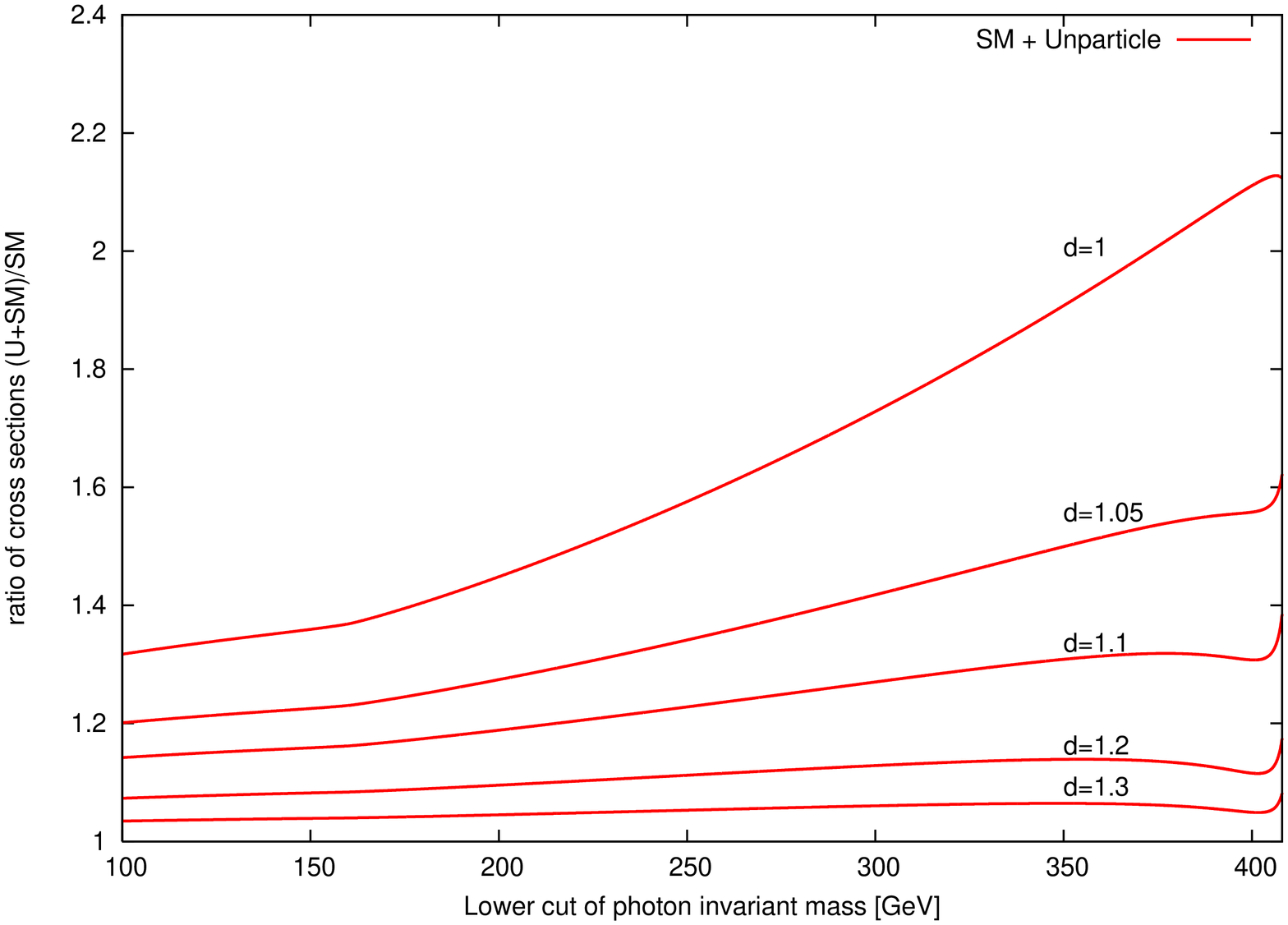}}
\end{center}
\caption{The same figure as Fig.~\ref{Fig7-2} 
but the beam polarizations are chosen as 
$(P_e, P_\ell,  P'_e, P'_\ell) = (\pm \pm \mp \mp)$.
}
\label{Fig10}
\end{figure}

\begin{figure}[p]
\begin{center}
\subfigure[The differential cross section]{\includegraphics[angle=-90, width=0.7\textwidth]{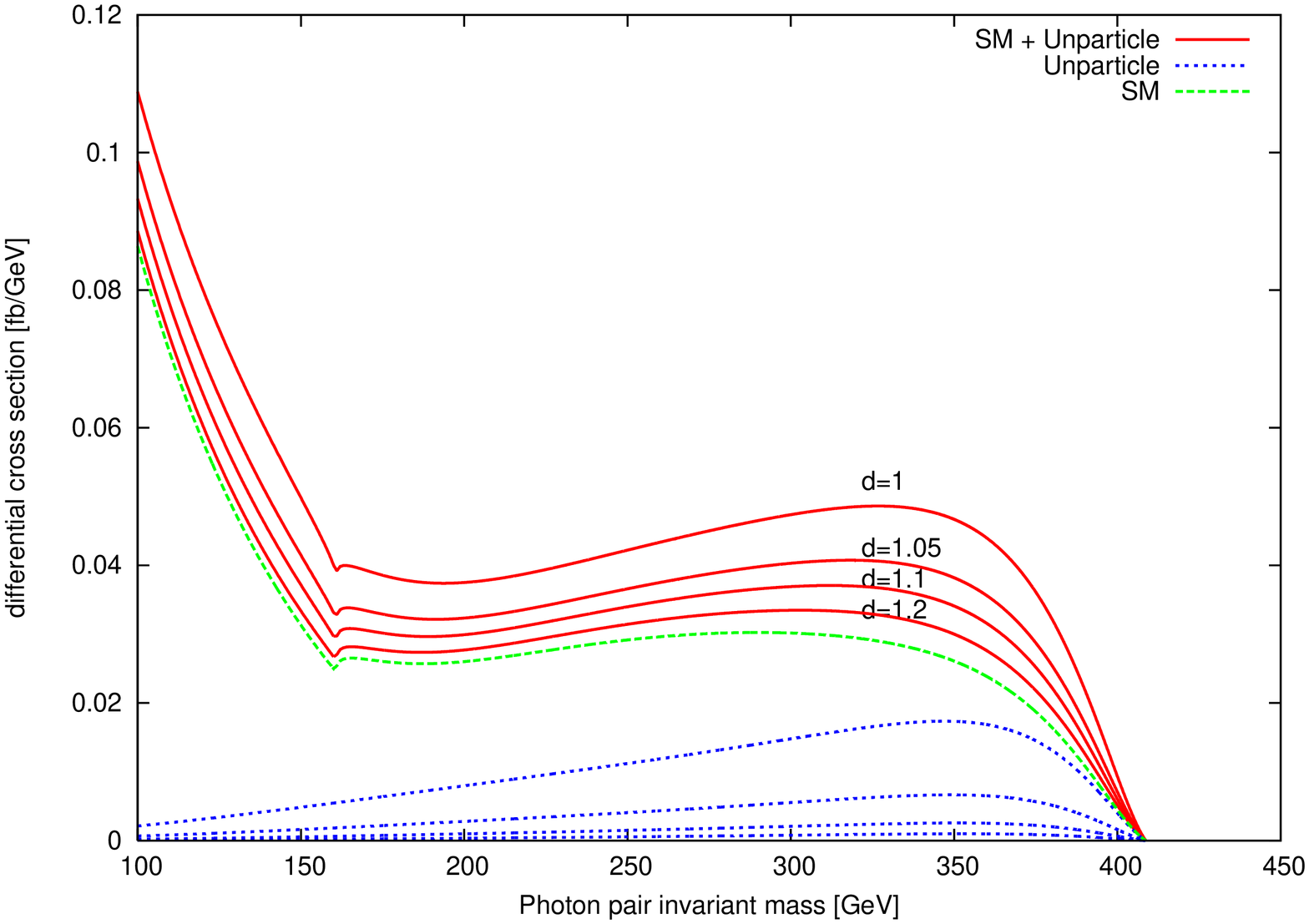}}
\subfigure[Ratio of cross section $\sigma_{{\cal U}+{\rm SM}}/\sigma_{\rm SM}$]{\includegraphics[angle=-90, width=0.7\textwidth]{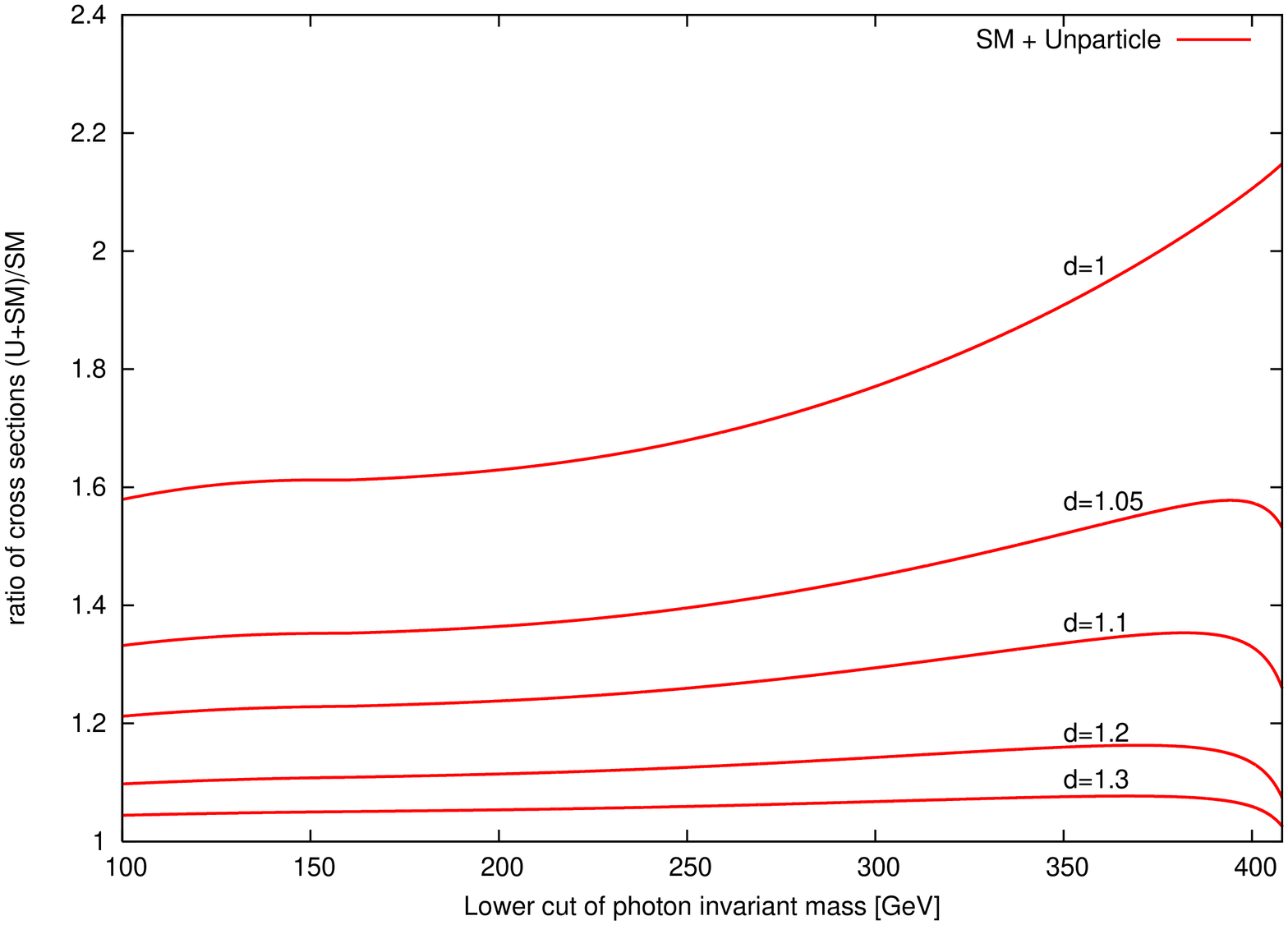}}
\end{center}
\caption{The same figure as Fig.~\ref{Fig7-2} 
but the beam polarizations are chosen as 
$(P_e, P_\ell,  P'_e, P'_\ell) = (\pm \pm \mp \pm)$ or $( \mp \pm \pm \pm)$.
}
\label{Fig11}
\end{figure}

\begin{figure}[p]
\begin{center}
\subfigure[The differential cross section]{\includegraphics[angle=-90, width=0.7\textwidth]{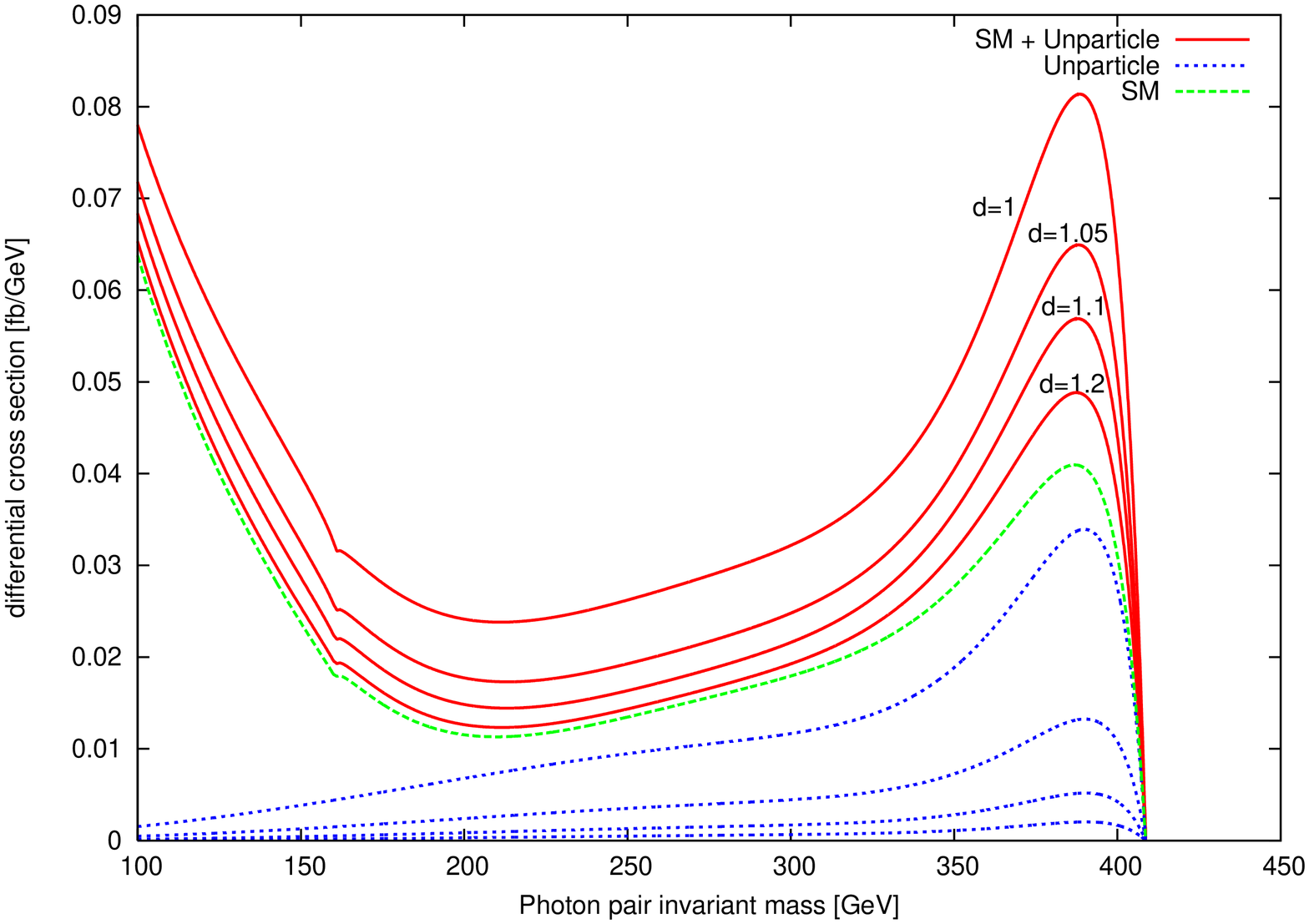}}
\subfigure[Ratio of cross section $\sigma_{{\cal U}+{\rm SM}}/\sigma_{\rm SM}$]{\includegraphics[angle=-90, width=0.7\textwidth]{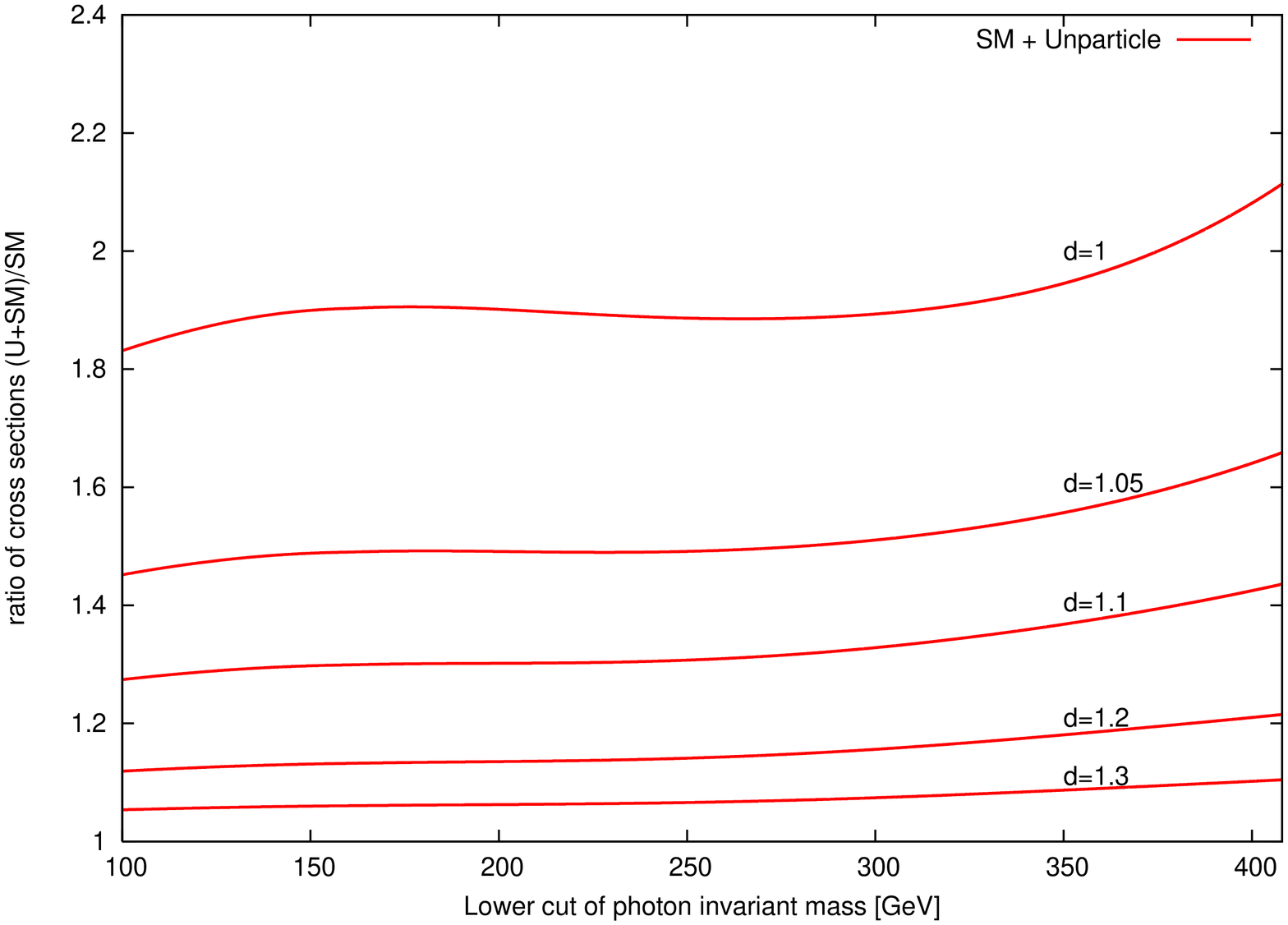}}
\end{center}
\caption{The same figure as Fig.~\ref{Fig7-2} 
but the beam polarizations are chosen as 
$(P_e, P_\ell,  P'_e, P'_\ell) = (\pm \mp \mp \pm)$.
}
\label{Fig12}
\end{figure}

\begin{figure}[p]
\begin{center}
\subfigure[The differential cross section]{\includegraphics[angle=-90, width=0.7\textwidth]{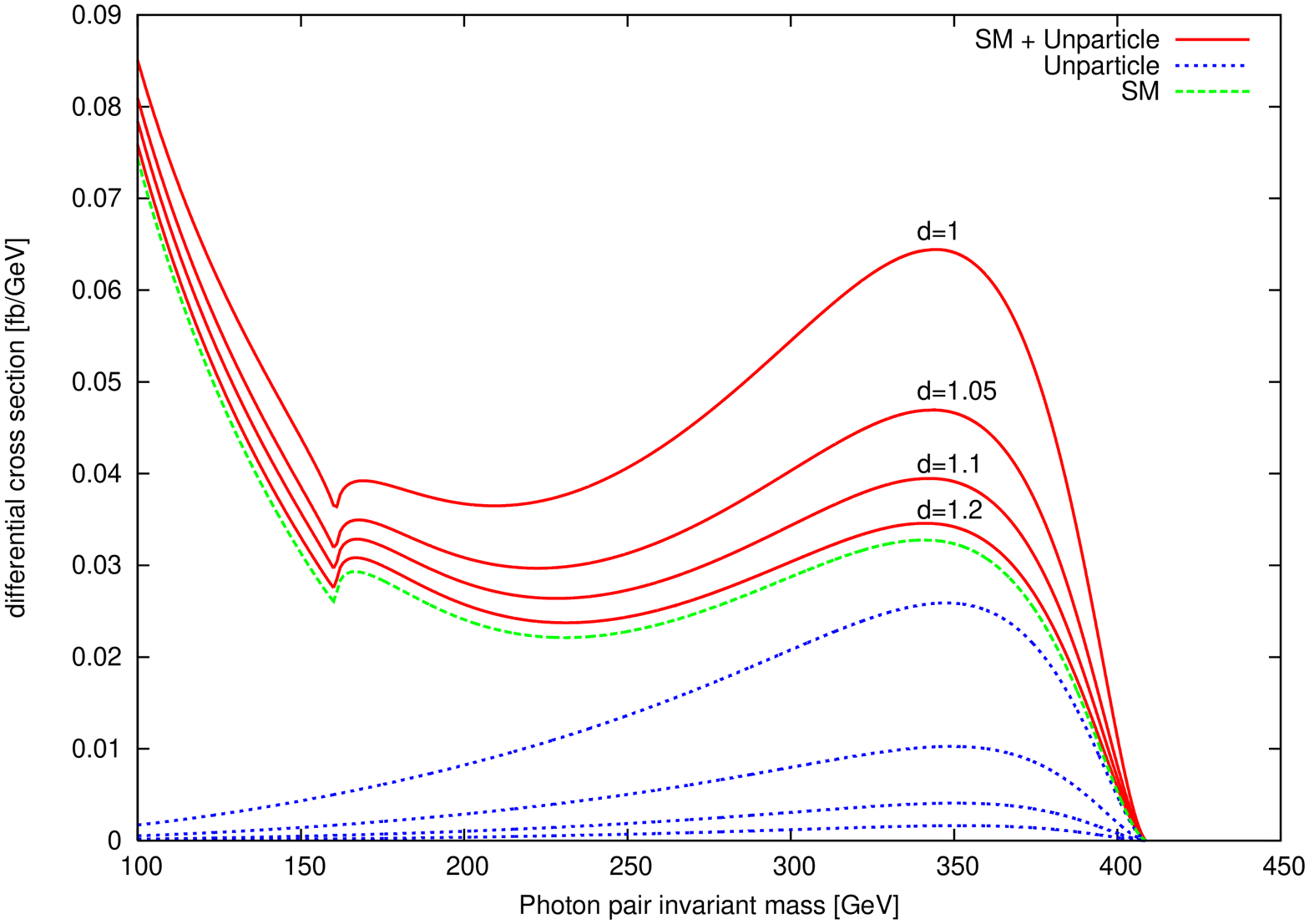}}
\subfigure[Ratio of cross section $\sigma_{{\cal U}+{\rm SM}}/\sigma_{\rm SM}$]{\includegraphics[angle=-90, width=0.7\textwidth]{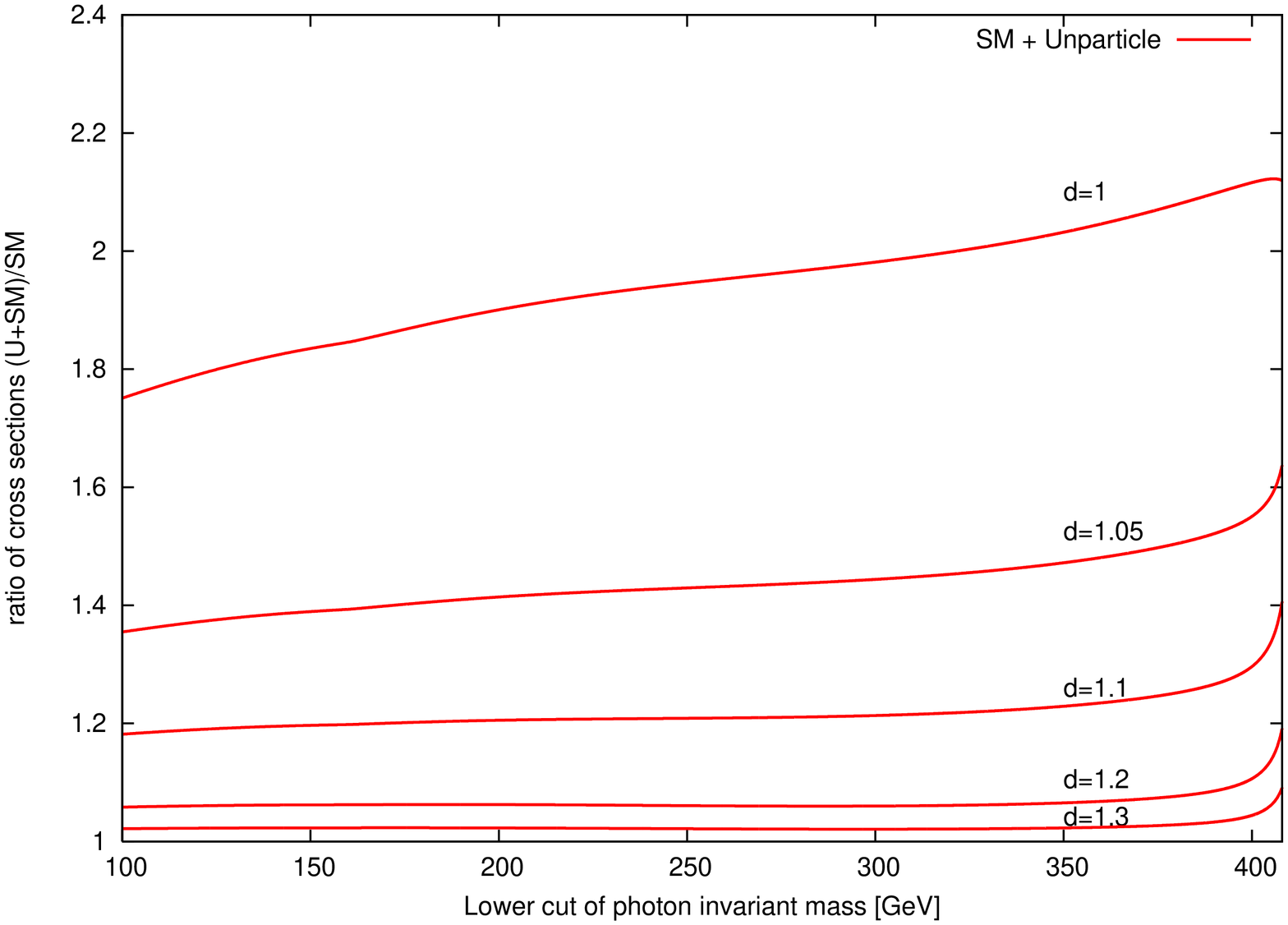}}
\end{center}
\caption{The same figure as Fig.~\ref{Fig7-2} 
but the beam polarizations are chosen as 
$(P_e, P_\ell,  P'_e, P'_\ell) = (\pm \pm \pm \mp)$ or $(\pm  \mp \pm \pm)$.
}
\label{Fig13}
\end{figure}

\begin{figure}[p]
\begin{center}
\subfigure[The differential cross section]{\includegraphics[angle=-90, width=0.7\textwidth]{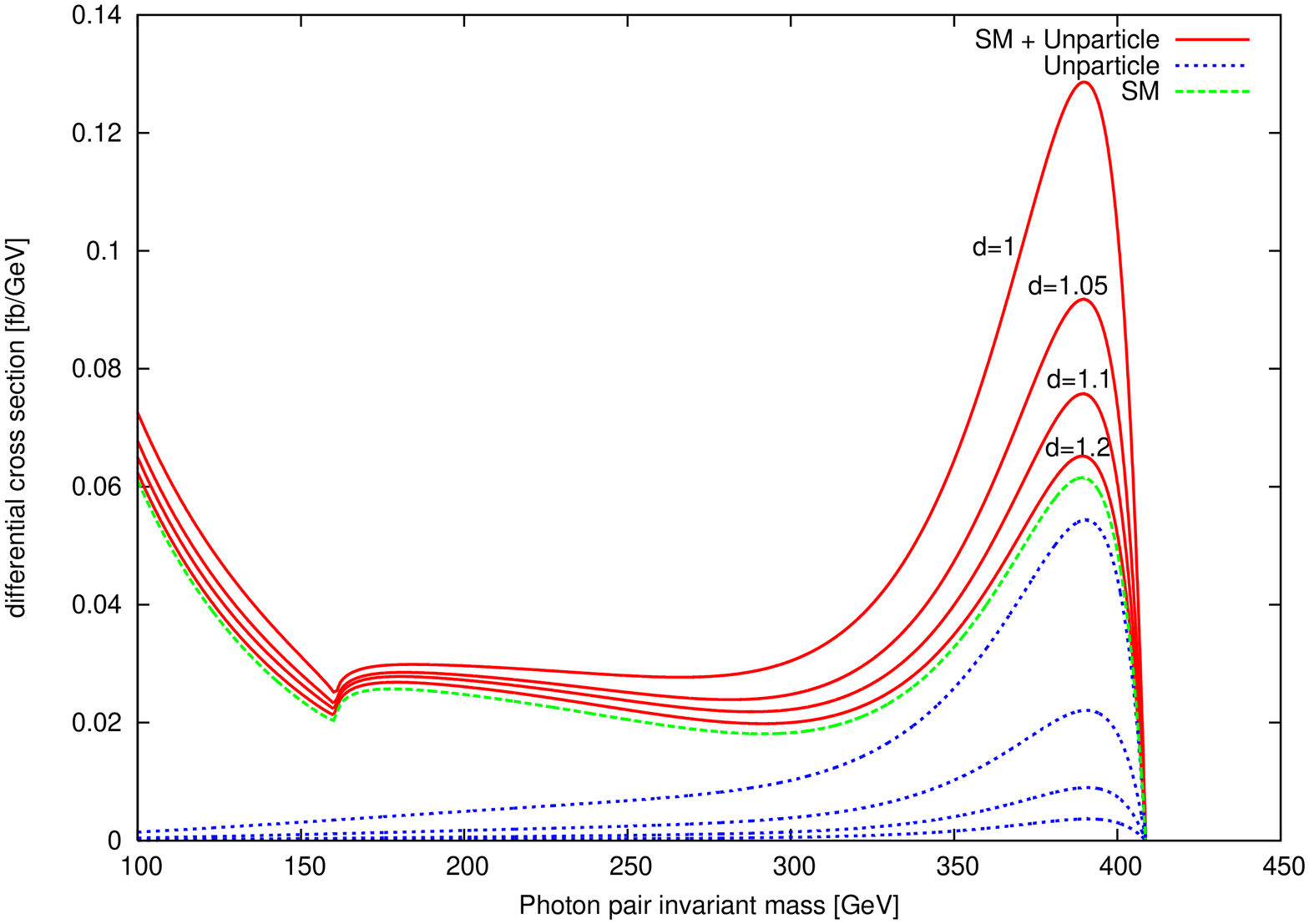}}
\subfigure[Ratio of cross section $\sigma_{{\cal U}+{\rm SM}}/\sigma_{\rm SM}$]{\includegraphics[angle=-90, width=0.7\textwidth]{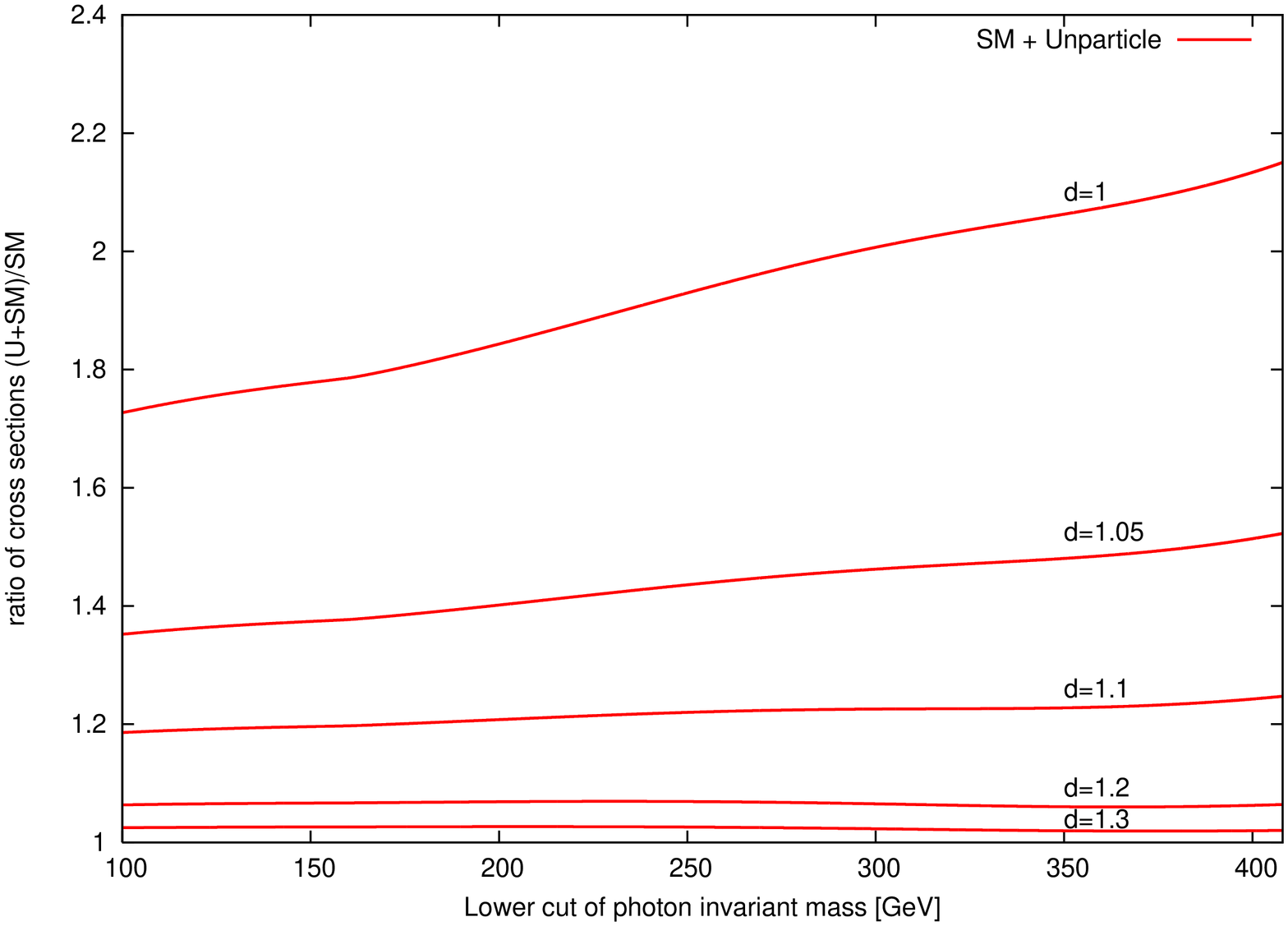}}
\end{center}
\caption{The same figure as Fig.~\ref{Fig7-2} 
but the beam polarizations are chosen as 
$(P_e, P_\ell,  P'_e, P'_\ell) = (\pm \mp \pm \mp)$.
}
\label{Fig14}
\end{figure}

The results using the polarized beams
 \footnote{
 In this paper, we assume the ideal case, $100 \%$ polarized beams. }
are shown in Figs.~\ref{Fig9}--\ref{Fig14}.
In these figures, it is shown for all combinations of polarizations in the beam.
It can be seen from these figures that the effects of the initial beam polarizations
can drastically change the behavior of the differential cross section $d\sigma/dM_{\gamma \gamma}$
as a function of the invariant mass of the final state photons.
We can enhance the signal over background ratio
to choose an appropriate initial beam polarizations.
For example, when we choose $(P_e, P_\ell,  P'_e, P'_\ell) = (+--+) ~{\rm or}~ (-++-)$,
the differential cross section $d\sigma/dM_{\gamma \gamma}$ has a sharp peak at high energy
as shown in Fig.~\ref{Fig12}.
Using these options, we can extract the information on $d_{\cal U}$.

\section{Summary}
We have considered the unparticle physics at the photon collider, 
 in particular, the unparticle effects on 
 the $\gamma \gamma \to \gamma \gamma$ process. 
Since this process occurs at loop level in the SM, 
 the unparticle effects can be significant 
 even if the cutoff scale is very high. 
We have analyzed the cross section for 
 the $\gamma \gamma \to \gamma \gamma$ process, 
 including the unparticle mediated process, and 
 found that even for $\Lambda=5$ TeV, 
 the unparticle effects cause the sizable deviations 
 from the SM results with the incident $e^+e^-$ collider energy 
 at $\sqrt{s}=500$ GeV. 
The dependence of the differential cross section 
of the final state photon invariant mass 
$d\sigma/dM_{\gamma \gamma}$
reflects the scaling dimension of the unparticle $d_{\cal U}$,  
therefore precision measurements of these dependences 
 may reveal the scaling dimension of the unparticle. 
It has also been shown that the effects of the initial beam polarizations
can drastically change the behaviors of the cross sections. 
 We can enhance the signal over background ratio 
in some magnitude by choosing the initial beam polarizations appropriate.

\section*{Acknowledgments}
The work of T.K. is supported by the Research Fellowship 
 of the Japan Society for the Promotion of Science (\#1911329).
The work of N.O. is supported in part by 
 the Grant-in-Aid for Scientific Research from the Ministry 
 of Education, Science and Culture of Japan (\#18740170). 
We would like to thank M.M. Nojiri and Y. Okada 
 for their useful comments and discussions. 

\section*{Note Added} 
After completing this work, 
 we became aware of a very recent paper \cite{U-photon0}, 
 in which the same subject on the unparticle effects 
 at the photon collider was studied. 
There exists a difference between their result and ours.
As pointed out in \cite{Chang:2008mk}, 
 the phase factor $e^{-i d_U \pi}$ associated 
 with the $s$-channel unparticle propagator 
 might not be taken care of properly in \cite{U-photon0} 
 while we treated it correctly and our result is consistent 
 with the one in \cite{Chang:2008mk}. 

\newpage

%
\pagestyle{empty}

\begin{thebibliography}{99}
\bibitem{U-photon2}
For a discussion to include the effect of a conformal symmetry breaking or the mass of the unparticle,
which could be induced from the VEV of the Higgs field, especially, in scope of the photon collider physics, please refer,
T.~Kikuchi, N.~Okada and M.~Takeuchi,
work in progress.
\end{thebibliography}
\end{document}